\documentclass[12pt]{article}

  \usepackage{graphicx}
  \usepackage{epsfig}
  \usepackage{amssymb}
  \usepackage{subfigure}
  \usepackage{axodraw}
  \usepackage{psfrag}
\evensidemargin - 1.truecm
\long\def\symbolfootnote[#1]#2{\begingroup%
\def\thefootnote{\fnsymbol{footnote}}\footnote[#1]{#2}\endgroup} 

\begin{document}

\newcommand{\be}{\begin{equation}}
\newcommand{\ee}{\end{equation}}
\newcommand{\nn}{\nonumber}
\newcommand{\bea}{\begin{eqnarray}}
\newcommand{\eea}{\end{eqnarray}}
\newcommand{\bfig}{\begin{figure}}
\newcommand{\efig}{\end{figure}}
\newcommand{\bc}{\begin{center}}
\newcommand{\ec}{\end{center}}
\newcommand{\bd}{\begin{displaymath}}
\newcommand{\ed}{\end{displaymath}}

\begin{titlepage}
\nopagebreak
{\flushright{
        \begin{minipage}{5cm}
        Freiburg-THEP 04/19\\
	UCLA/04/TEP/60\\
        {\tt hep-ph/0411321}\\
        \end{minipage}        }

}
\vspace*{-1.5cm}                        
\vskip 2.5cm
\begin{center}
\boldmath
{\Large \bf Two-Loop  $N_{F}=1$  QED Bhabha Scattering: \\ [1mm] 
Soft Emission and Numerical Evaluation \\ [3mm]
of the Differential Cross-section}\unboldmath
\vskip 1.2cm
{\large  R.~Bonciani$\rm \, ^{a, \,}$\symbolfootnote[1]{Email: 
{\tt Roberto.Bonciani@physik.uni-freiburg.de}}}
{\large A.~Ferroglia$\rm \, ^{a, \,}$\symbolfootnote[2]{Email: 
{\tt Andrea.Ferroglia@physik.uni-freiburg.de}}},
{\large P.~Mastrolia$\rm \, ^{b, \,}$\symbolfootnote[3]{Email: 
{\tt mastrolia@physics.ucla.edu}}}, \\[2mm] 
{\large E.~Remiddi$\rm \, ^{c, \, d, \,}$\symbolfootnote[4]{Email: 
{\tt Ettore.Remiddi@bo.infn.it}}},
and  {\large  J. J. van der Bij$\rm \, ^{a, \,}$\symbolfootnote[5]{Email: 
{\tt jochum@physik.uni-freiburg.de}}}
\vskip .7cm
{\it $\rm ^a$ Fakult\"at f\"ur Mathematik und Physik, 
Albert-Ludwigs-Universit\"at
Freiburg, \\ D-79104 Freiburg, Germany} 
\vskip .3cm
{\it $\rm ^b$ Department of Physics and Astronomy, UCLA,
Los Angeles, CA 90095-1547} 
\vskip .3cm
{\it $\rm ^c$ Theory Division, CERN, CH-1211 Geneva 23, Switzerland} 
\vskip .3cm
{\it $\rm ^d$ Dipartimento di Fisica dell'Universit\`a di Bologna, and
INFN, Sezione di Bologna, I-40126 Bologna, Italy} 
\end{center}
\vskip .5cm


\begin{abstract}  

Recently, 
we evaluated the virtual cross-section for Bhabha scattering in pure QED, 
up to corrections of order $\alpha^4 (N_F =1)$. This calculation is 
valid for arbitrary values of the squared center of mass  energy $s$ and momentum transfer 
$t$; the electron and positron mass $m$ was considered a finite, non vanishing 
quantity. In the present work, we supplement the previous calculation by 
considering the contribution of the soft photon emission diagrams to the 
differential cross-section, up to and including terms of order  $\alpha^4 (N_F
=1)$. Adding the contribution of the real corrections to the renormalized 
virtual ones, we obtain an UV and IR finite differential cross-section; we
evaluate this quantity numerically for a significant set of values of the
squared center of mass  energy $s$.

\vskip .3cm 
\flushright{
        \begin{minipage}{12.3cm}
{\it Key words}: Feynman
diagrams, Multi-loop calculations,  Box diagrams, \hspace*{2cm} Bhabha 
scattering \\
{\it PACS}: 11.15.Bt, 12.20.Ds
        \end{minipage}        }
\end{abstract}
\vfill
\end{titlepage}    


\section{Introduction \label{Intro}}

The relevance of the Bhabha scattering process ($e^+  e^- \rightarrow 
e^+  e^-$) in the study of the phenomenology of particle physics can hardly be
overestimated. This is due to the fact that Bhabha scattering is the process
employed to determine the luminosity of all the present $e^+ - e^-$ colliders, 
at both high ($\sim 100 \, \mbox{GeV}$) and intermediate ($\sim 1 - 10 \,
\mbox{GeV}$) energies, as well as in a future linear collider. For this reason, 
Bhabha scattering has been studied in great detail within the context of
the electroweak Standard Model (we refer the interested reader to \cite{reviews} 
and references therein).  
  
In recent years, it was pointed out that, while the one-loop radiative 
corrections to Bhabha scattering and the corresponding real emission
corrections  were well known within the full electroweak Standard Model 
\cite{Bhabha1loopC,Bhabha1loopB}, the two-loop corrections had not been 
calculated even in pure QED, although a large amount of work was devoted to
the  study of the contributions enhanced by factors of $\ln(s/m_e^2)$
\cite{TUTTI}. The reason for this situation was identified in the 
technical problem  of calculating the necessary two-loop box diagrams. Several
groups started to work on different aspects of the problem. In \cite{Bern},
the  two-loop QED  virtual cross-section in the limit of zero electron mass was
calculated, while   in \cite{Bas} the IR divergent structure of that result was
carefully studied.  A very interesting and ambitious project aiming at the
complete evaluation of  the Bhabha scattering cross-section in QED, without
neglecting the electron mass  $m$, was presented in \cite{Czakon}. Some of the
necessary Feynman diagrams were  calculated in \cite{RoPieRem1,us,Gudrum}. 

By employing the results of  \cite{RoPieRem1,us}, it was possible to complete
the calculation of the virtual Bhabha scattering unpolarized  differential
cross-section, including the contribution of two-loop graphs involving a closed
fermion loop (conventionally indicated as corrections of order  $\alpha^4 (N_F
=1)$, where $\alpha$ is the fine structure constant) \cite{us2}. The cross 
section presented in \cite{us2} is valid for arbitrary values of the squared 
center of mass energy $s$ and momentum transfer $t$. The full dependence of the 
cross-section on the electron and positron mass $m$ was retained. In calculating 
the Feynman diagrams, both UV and IR divergences were regularized with the 
continuous $D$-dimensional regularization scheme \cite{DimReg}, while the 
diagrams were evaluated analytically in \cite{RoPieRem1, us} by means of the 
Laporta algorithm \cite{LapAlg} and the differential equations technique for the
evaluation of the master integrals \cite{DiffEq}. In
\cite{us2}, the renormalization program was carried out in the on-shell scheme;
therefore the final result, expressed in terms of 1- and 2-dimensional harmonic
polylogarithms (HPLs, 2dHPLs) \cite{HPLs} of maximum weight 3, is free from UV
divergences. However, the cross-section of \cite{us2} still includes
IR divergent terms that appear as a pole in $(D-4)$.

As is well known, the IR divergent term in the cross-section of \cite{us2}
cancels if one adds to it the contribution of events of the type $e^+  e^-
\rightarrow e^+  e^- \gamma$, in the limit in which the photon in the final
state carries an energy which is small with respect to the squared center of mass 
energy. Such events are commonly referred to as {\em{soft}} photon events, and
their contributions to the cross-section are known as
{\em real} (as opposed to virtual) corrections.

The purpose of the present paper is to calculate the contribution of the the
soft radiation diagrams (up to and including order $\alpha^4 (N_F =1)$) to the 
Bhabha scattering differential cross-section, in order to verify that the IR
divergent terms present in this contribution cancel against the IR
divergent terms in the cross-section in \cite{us2}, as well as to ultimately
obtain the IR finite differential cross-section by adding the virtual and real
corrections.

In calculating the contribution of the soft photon emission diagrams to the
cross-section, we integrated over the soft photon phase space,  taking into
consideration  the emission of photons with an energy smaller than a certain
energy threshold $\omega$. This threshold is supposed to be small  with respect
to the beam energy $E$. In other words, we assumed that the idealized detector,
to be used to measure the differential cross-section we calculated, can tag all
the events in which one photon appears in the final state, provided that the
photon carries an energy larger than the chosen threshold.  We  also assumed
that all the events in which a photon with an energy smaller than the threshold
appears in the final state, are for the idealized detector indistinguishable 
from the Bhabha scattering events without photon emission. This is the standard
textbook approach to the calculation of the real corrections to a given
cross-section; admittedly, it is not enough to consider cross-sections
measured in realistic experiments, where  other aspects must be taken into
account (e.~g. hard bremsstrahlung effects, detector geometry). Very often, the
experimental set up is so complex that the  only effective tool to obtain a
realistic cross-section is the Monte Carlo method.  Nevertheless, we thought it
useful to calculate the soft photon emission following this approach, in order
to  diagrammatically show how the cancellation of the IR divergences works,
as well as to provide a benchmark for future, more realistic numerical 
calculations. 

Instead of  presenting the lengthy analytical expression of the IR finite
differential cross-section, we preferred to implement computer codes that
evaluate the differential cross-section at order $\alpha^4 (N_F =1)$
for arbitrary values of the beam energy and scattering angle
in the center of mass, $E$ and $\theta$, respectively.   

We have found that the corrections at order $\alpha^4 (N_F =1)$ are positive
and  very small with respect to the corrections of order $\alpha^3$ (which are
negative), so that their constructive contribution is strongly suppressed in
the energy range of interest at present and future colliders. The relative
weight of the corrections of order $\alpha^3$ and  $\alpha^4 (N_F =1)$ does
increase in magnitude with the beam energy and, at a given energy, with the
scattering angle.  In order to check our numerical consistency, we reproduced
the  results in \cite{Bhabha1loopC} and \cite{Bhabha1loopB} revelant for our
purposes. 

Our routines for the numerical evaluation of  the Bhabha scattering
cross-section up to corrections of order $\alpha^4 (N_F =1)$,  written
both in {\tt Mathematica} \cite{Mathematica} and  in {\tt Fortran77}, can be
obtained from the authors \cite{BHABHAPAGE}. 

The paper is organized as follows: in Section~\ref{Real}, we discuss the
calculation of the soft real corrections to the Bhabha scattering  differential
cross-section \footnote{Here and in the following, we always consider 
differential cross-sections summed over the spins of the final state particles
and averaged over the spin of the incoming  electron and positron.}, up to and
including terms of order  $\alpha^4 (N_F =1)$.  In addition, we investigate the
cancellation of the IR divergent terms  of the Bhabha scattering virtual
cross-section calculated in \cite{us2}  against the IR divergent terms
originating from the soft photon emission  diagrams considered here. In
Section~\ref{Num}, we present the numerical  results obtained by evaluating the
IR finite Bhabha scattering cross-section  (given by the sum of virtual and
soft corrections up to terms of order  $\alpha^4 (N_F =1)$), for different
significant choices of the  beam energy $E$. Also, we compare the complete
cross-section with its expansion in powers of the electron mass. We find that
the first term in the expansion fits to sufficient approximation the numerical
value of the complete cross-section for all the beam energies relevant in
present and future colliders. In Section~\ref{Conc}, we present our
conclusions. Finally, two appendices include the expression of the integrals
occurring in the evaluation of the soft radiative corrections relevant to our
calculation, as well as the leading terms in the expansion  in the limit
 $m \to 0$ of the Bhabha scattering differential cross-section we studied. 
 

\section{The Real Corrections \label{Real}}

In this Section, we discuss the calculation of the real corrections due to the
emission of a soft photon to the Bhabha scattering unpolarized
differential cross-section in pure QED. In particular, we obtain the real
corrections of order $\alpha^3$ and  $\alpha^4 (N_F =1)$. In both cases, we
must consider events involving a single soft photon in the final state:
\be
e^-\left(p_1 \right) + e^+\left(p_2 \right) \longrightarrow
e^-\left(p_3 \right) + e^+\left(p_4 \right) + \gamma\left(k \right) \, ,
\ee
where $p_1$, $p_2$, $p_3$, $p_4$, and $k$ are the momenta carried by the
incoming electron, incoming  positron, outgoing electron, outgoing positron,
and outgoing soft photon, respectively.
All of the particles in the initial and final states are on-shell, so that
$p_i^2 = -m^2$ ($i =1,4$) and $k^2 =0$. 
Therefore, we introduce  the quantities $s$ and $s'$ defined as follows:
\be
s = - (p_1 + p_2)^2 \, , \quad s' = - (p_3 + p_4)^2 \, .
\ee 
By definition, the soft photon approximation consists of neglecting $k$ 
in the numerator of the scattering amplitude and of setting $s' = s$ everywhere.
In this approximation, the kinematical relations that link the Mandelstam
invariants $s$, $t$, and $u$ to the beam energy ($E$) and to the scattering angle
in the center of mass frame ($\theta$), are
\bea
s &=& -P^2 \equiv - (p_1 + p_2)^2 \approx - (p_3+p_4)^2 = 4 E^2 \, ,\\ 
t &=& -Q^2 \equiv - (p_1 - p_3)^2 \approx - (p_2-p_4)^2 = -4 \left(E^2
-m^2\right)\sin^2{\frac{\theta}{2}} \, ,\\ 
u &=& -V^2 \equiv - (p_1 - p_4)^2 \approx - (p_2-p_3)^2 = -4 \left(E^2
-m^2\right)\cos^2{\frac{\theta}{2}} \,,
\eea
with
\be
s + t + u = 4 m^2 \, .
\ee
In the following, we often employ the dimensionless variables $x$, $y$, and $z$,
related to the Mandelstam invariants $s$, $t$, and $u$, by the relations
\bea
s &=&  m^2 \frac{(1+x)^2}{x} \, , \quad x = \frac{\sqrt{s} - \sqrt{s- 4
m^2}}{\sqrt{s} + \sqrt{s - 4m^2}} \, , \label{eqx}\\
t &=& -m^2 \frac{(1-y)^2}{y} \, , \qquad y = \frac{\sqrt{4 m^2 - t} - 
\sqrt{-t}}{\sqrt{4 m^2 - t} + \sqrt{-t}} \, ,\\
u &=& -m^2 \frac{(1-z)^2}{z} \, , \qquad z = \frac{\sqrt{4 m^2 - u} - 
\sqrt{-u}}{\sqrt{4 m^2 - u} + \sqrt{-u}} \, , \label{eqz}\\
\eea
which are valid in the physical region $s \ge 4 m^2$, $t,u \le 0$.

The complete Bhabha scattering differential cross-section up to order 
$\alpha^4 (N_F=1)$ can be written as follows:
\be
\frac{d \sigma^{T}(s,t,m^2)}{d \Omega} = 
\frac{d \sigma_0(s,t,m^2)}{d \Omega} + \left( \frac{\alpha}{\pi} \right) 
\frac{d \sigma^{T}_1(s,t,m^2)}{d \Omega} + 
\left( \frac{\alpha}{\pi} \right)^2
\frac{d \sigma^{T}_2(s,t,m^2)}{d \Omega} \, ,
\ee
where $\sigma_0(s,t,m^2)$ is the tree-level (Born) cross-section
\bea
\frac{d \sigma_0(s,t,m^2)}{d \Omega} &=& \frac{\alpha^2}{s} \Bigl\{
\frac{1}{s^2}\left[s t + \frac{s^2}{2} + \left(t - 2 m^2 \right)^2 \right]+
\frac{1}{t^2}\left[s t + \frac{t^2}{2} + \left(s - 2 m^2 \right)^2 \right]
\nn \\ & & + \frac{1}{s t}\left[\left(s+t\right)^2 - 4 m^4\right] 
\Bigr\}\, ,
\label{BORN}
\eea
and $\sigma_i^{T}(s,t,m^2)$ ($i=1,2$) are the sum of the virtual and real
corrections at order $\alpha^3$ and $\alpha^4 (N_F=1)$, respectively.

\boldmath
\subsection{Real Corrections at Order $\alpha^3$}
\unboldmath

We first consider the order $\alpha^3$ contribution to the cross-section,
which is given by:
\be
\left( \frac{\alpha}{\pi} \right)\frac{d \sigma_1^{T}(s,t,m^2)}{d \Omega} = 
 \left( \frac{\alpha}{\pi} \right) 
 \left[ \frac{d \sigma^{V}_1(s,t,m^2)}{d \Omega}
+ \frac{d \sigma^{S}_1(s,t,m^2)}{d \Omega} \right] \, ,
\label{1LOOP}
\ee
where the superscripts $V$ and $S$ stand for ``virtual'' and ``soft''.

The one-loop virtual cross-section $d \sigma^{V}_1/d \Omega$ can be found in
Eq.~(67) of \cite{us2}; we devote the remaining part of this subsection to the
calculation of $d \sigma^{S}_1/d \Omega$.

\begin{figure}
\vspace*{.6cm}
\[\vcenter{
\hbox{
  \begin{picture}(0,0)(0,0)
\SetScale{.8}
  \SetWidth{.5}
\ArrowLine(0,20)(30,50)
\ArrowLine(-30,50)(-15,35)
\ArrowLine(-15,35)(0,20)
\ArrowLine(30,-50)(0,-20)
\ArrowLine(0,-20)(-30,-50)
\Photon(0,-20)(0,20){2}{6}
\Photon(-15,35)(0,50){2}{4}
\LongArrow(-30,37)(-10,17)
\LongArrow(-30,-37)(-10,-17)
\LongArrow(10,17)(30,37)
\LongArrow(10,-17)(30,-37)
\LongArrow(-7,35)(3,45)
\Text(0,-20)[cb]{{\footnotesize (a)}}
\Text(-7,4)[cb]{{\footnotesize $p_1$}}
\Text(7.2,4)[cb]{{\footnotesize $p_3$}}
\Text(-7,-5)[cb]{{\footnotesize $p_2$}}
\Text(7.2,-5)[cb]{{\footnotesize $p_4$}}
\Text(3,12)[cb]{{\footnotesize $k$}}
\end{picture}}  
}
\hspace{3cm}
  \vcenter{
\hbox{
  \begin{picture}(0,0)(0,0)
\SetScale{.8}
  \SetWidth{.5}
\ArrowLine(0,20)(15,35)
\ArrowLine(15,35)(30,50)
\ArrowLine(-30,50)(0,20)
\ArrowLine(30,-50)(0,-20)
\ArrowLine(0,-20)(-30,-50)
\Photon(0,-20)(0,20){2}{6}
\Photon(15,35)(30,20){2}{4}
\Text(0,-20)[cb]{{\footnotesize (b)}}
\end{picture}}  
}
\hspace{3cm}
  \vcenter{
\hbox{
  \begin{picture}(0,0)(0,0)
\SetScale{.8}
  \SetWidth{.5}
\ArrowLine(0,20)(30,50)
\ArrowLine(-30,50)(0,20)
\ArrowLine(-15,-35)(-30,-50)
\ArrowLine(0,-20)(-15,-35)
\ArrowLine(30,-50)(0,-20)
\Photon(0,-20)(0,20){2}{6}
\Photon(-15,-35)(0,-50){2}{4}
\Text(0,-20)[cb]{{\footnotesize (c)}}
\end{picture}}  
}
\hspace{3cm}
  \vcenter{
\hbox{
  \begin{picture}(0,0)(0,0)
\SetScale{.8}
  \SetWidth{.5}
\ArrowLine(0,20)(30,50)
\ArrowLine(-30,50)(0,20)
\ArrowLine(0,-20)(-30,-50)
\ArrowLine(30,-50)(15,-35)
\ArrowLine(15,-35)(0,-20)
\Photon(0,-20)(0,20){2}{6}
\Photon(15,-35)(30,-20){2}{4}
\Text(0,-20)[cb]{{\footnotesize (d)}}
\end{picture}}  
}
\]
\vspace*{2cm}
\[\vcenter{
\hbox{
  \begin{picture}(0,0)(0,0)
\SetScale{.8}
  \SetWidth{.5}
\ArrowLine(-45,30)(-30,15)
\ArrowLine(-30,15)(-15,0)
\ArrowLine(-15,0)(-45,-30)
\ArrowLine(15,0)(45,30)
\ArrowLine(45,-30)(15,0)
\Photon(-15,0)(15,0){2}{5}
\Photon(-30,15)(-15,30){2}{4}
\Text(0,-12)[cb]{{\footnotesize (e)}}
\end{picture}}  
}
\hspace{3cm}
  \vcenter{
\hbox{
  \begin{picture}(0,0)(0,0)
\SetScale{.8}
  \SetWidth{.5}
\ArrowLine(-45,30)(-15,0)
\ArrowLine(-15,0)(-30,-15)
\ArrowLine(-30,-15)(-45,-30)
\ArrowLine(15,0)(45,30)
\ArrowLine(45,-30)(15,0)
\Photon(-15,0)(15,0){2}{5}
\Photon(-30,-15)(-15,-30){2}{4}
\Text(0,-12)[cb]{{\footnotesize (f)}}
\end{picture}}  
}
\hspace{3cm}
  \vcenter{
\hbox{
  \begin{picture}(0,0)(0,0)
\SetScale{.8}
  \SetWidth{.5}
\ArrowLine(-45,30)(-15,0)
\ArrowLine(-15,0)(-45,-30)
\ArrowLine(15,0)(30,15)
\ArrowLine(30,15)(45,30)
\ArrowLine(45,-30)(15,0)
\Photon(-15,0)(15,0){2}{5}
\Photon(30,15)(45,0){2}{4}
\Text(0,-12)[cb]{{\footnotesize (g)}}
\end{picture}}  
}
\hspace{3cm}
  \vcenter{
\hbox{
  \begin{picture}(0,0)(0,0)
\SetScale{.8}
  \SetWidth{.5}
\ArrowLine(-45,30)(-15,0)
\ArrowLine(-15,0)(-45,-30)
\ArrowLine(15,0)(45,30)
\ArrowLine(45,-30)(30,-15)
\ArrowLine(30,-15)(15,0)
\Photon(-15,0)(15,0){2}{5}
\Photon(30,-15)(45,0){2}{4}
\Text(0,-12)[cb]{{\footnotesize (h)}}
\end{picture}}  
}
\]
\vspace*{.6cm}
\caption[]{\it Diagrams contributing to the real corrections at order $\alpha^3$.}
\label{tlph}
\end{figure}
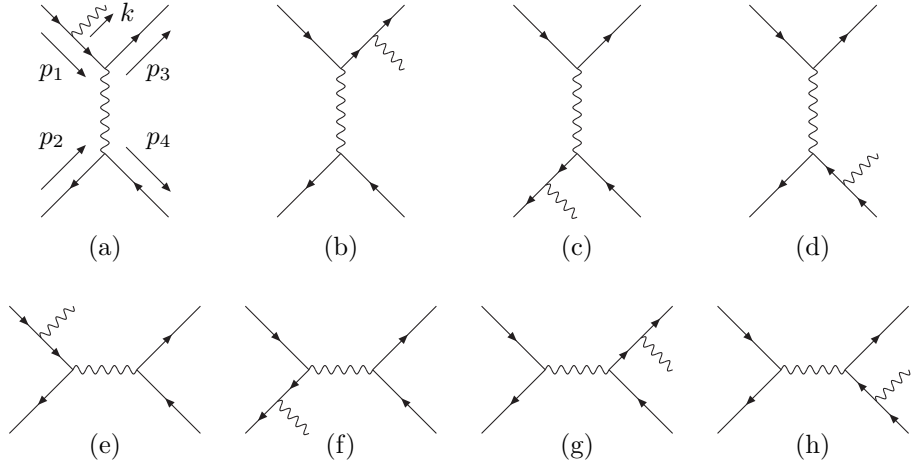

The diagrams contributing to the real corrections to the Bhabha scattering
cross-section at order $\alpha^3$ are shown in Fig.~\ref{tlph}; the real photon
can be emitted by any of the incoming or outgoing fermion lines of the $s$- and
$t$-channel Bhabha scattering tree-level diagrams.

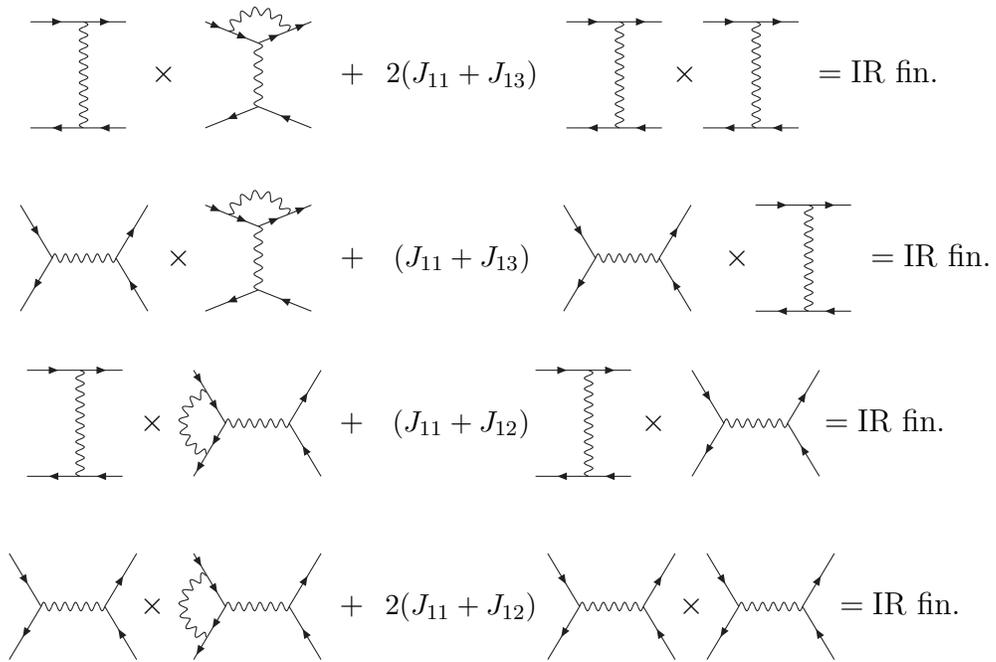
\begin{figure}
\vspace*{1cm}
\[
\hspace*{1cm}
\vcenter{
\hbox{
  \begin{picture}(0,0)(0,0)
\SetScale{.8}
  \SetWidth{.5}
\ArrowLine(-50,25)(-25,25)
\ArrowLine(-25,25)(-5,25)
\ArrowLine(-25,-25)(-50,-25)
\ArrowLine(-5,-25)(-25,-25)
\Photon(-25,25)(-25,-25){2}{10}
\end{picture}}  
}
\hspace{0.1cm}
\times
\hspace{1.cm}
\vcenter{\hbox{\begin{picture}(0,0)(0,0)
  \SetScale{0.8}
  \SetWidth{.5}
\ArrowLine(-25,25)(-15,21) 
\ArrowLine(-15,21)(0,15) 
\ArrowLine(15,21)(25,25) 
\ArrowLine(0,15)(15,21) 
\ArrowLine(0,-15)(-25,-25)
\ArrowLine(25,-25)(0,-15)
\Photon(0,15)(0,-15){2}{5}
\PhotonArc(0,15)(15,22.5,157.5){2}{6}
\end{picture}}}
\hspace{1cm}
+
\hspace{4cm}
  \vcenter{
\hbox{\begin{picture}(0,0)(0,0)
\SetScale{.8}
  \SetWidth{.5}
\ArrowLine(-50,25)(-25,25)
\ArrowLine(-25,25)(-5,25)
\ArrowLine(-25,-25)(-50,-25)
\ArrowLine(-5,-25)(-25,-25)
\Photon(-25,25)(-25,-25){2}{10}
\Text(-28,-2)[cb]{{\small $2 (J_{11} + J_{13}) $}}
\end{picture}}
}
\hspace{-0.1cm}
\times
\hspace{1.4cm}
  \vcenter{
\hbox{\begin{picture}(0,0)(0,0)
\SetScale{.8}
  \SetWidth{.5}
\ArrowLine(-50,25)(-25,25)
\ArrowLine(-25,25)(-5,25)
\ArrowLine(-25,-25)(-50,-25)
\ArrowLine(-5,-25)(-25,-25)
\Photon(-25,25)(-25,-25){2}{10}
\end{picture}}
}
=
\mbox{IR fin.}
\]
\vspace*{1cm}
\[
\hspace*{1cm}
\vcenter{
\hbox{
  \begin{picture}(0,0)(0,0)
\SetScale{.8}
  \SetWidth{.5}
\ArrowLine(-30,25)(-15,0) 
\ArrowLine(-15,0)(-30,-25) 
\ArrowLine(15,0)(30,25) 
\ArrowLine(30,-25)(15,0) 
\Photon(-15,0)(15,0){2}{6}
\end{picture}}  
}
\hspace{1cm}
\times
\hspace{.8cm}
\vcenter{\hbox{\begin{picture}(0,0)(0,0)
  \SetScale{0.8}
  \SetWidth{.5}
\ArrowLine(-25,25)(-15,21) 
\ArrowLine(-15,21)(0,15) 
\ArrowLine(15,21)(25,25)  
\ArrowLine(0,15)(15,21) 
\ArrowLine(0,-15)(-25,-25)
\ArrowLine(25,-25)(0,-15)
\Photon(0,15)(0,-15){2}{5}
\PhotonArc(0,15)(15,22.5,157.5){2}{6}
\end{picture}}}
\hspace{1cm}
+
\hspace{3.4cm}
  \vcenter{
\hbox{\begin{picture}(0,0)(0,0)
\SetScale{.8}
  \SetWidth{.5}
\ArrowLine(-30,25)(-15,0) 
\ArrowLine(-15,0)(-30,-25) 
\ArrowLine(15,0)(30,25) 
\ArrowLine(30,-25)(15,0) 
\Photon(-15,0)(15,0){2}{6}
\Text(-22,-2)[cb]{{\small $( J_{11} + J_{13})$}}
\end{picture}}
}
\hspace{1.2cm}
\times
\hspace{1.4cm}
  \vcenter{
\hbox{\begin{picture}(0,0)(0,0)
\SetScale{.8}
  \SetWidth{.5}
\ArrowLine(-50,25)(-25,25)
\ArrowLine(-25,25)(-5,25)
\ArrowLine(-25,-25)(-50,-25)
\ArrowLine(-5,-25)(-25,-25)
\Photon(-25,25)(-25,-25){2}{10}
\end{picture}}
}
=
\mbox{IR fin.}
\]

\vspace*{1cm}
\[
\hspace*{1.04cm}
\vcenter{
\hbox{
  \begin{picture}(0,0)(0,0)
\SetScale{.8}
  \SetWidth{.5}
\ArrowLine(-50,25)(-25,25)
\ArrowLine(-25,25)(-5,25)
\ArrowLine(-25,-25)(-50,-25)
\ArrowLine(-5,-25)(-25,-25)
\Photon(-25,25)(-25,-25){2}{10}
\end{picture}}  
}
\hspace*{0.cm}
\times
\hspace*{1.cm}
\vcenter{\hbox{
  \begin{picture}(0,0)(0,0)
\SetScale{.8}
  \SetWidth{.5}
\ArrowLine(-30,25)(-25,16.67) 
\ArrowLine(-25,16.67)(-15,0) 
\ArrowLine(-15,0)(-25,-16.67) 
\ArrowLine(-25,-16.67)(-30,-25) 
\ArrowLine(15,0)(30,25) 
\ArrowLine(30,-25)(15,0) 
\Photon(-15,0)(15,0){2}{6}
\PhotonArc(-20,0)(15,106,254){2}{7}
\end{picture}}  }
\hspace*{1cm}
+
\hspace*{3.6cm}
  \vcenter{
\hbox{\begin{picture}(0,0)(0,0)
\SetScale{.8}
  \SetWidth{.5}
\ArrowLine(-50,25)(-25,25)
\ArrowLine(-25,25)(-5,25)
\ArrowLine(-25,-25)(-50,-25)
\ArrowLine(-5,-25)(-25,-25)
\Photon(-25,25)(-25,-25){2}{10}
\Text(-24,-2)[cb]{{\small $( J_{11} + J_{12}) $}}
\end{picture}}
}
\hspace{-0.1cm}
\times
\hspace{1.1cm}
  \vcenter{
\hbox{\begin{picture}(0,0)(0,0)
\SetScale{.8}
  \SetWidth{.5}
\ArrowLine(-30,25)(-15,0) 
\ArrowLine(-15,0)(-30,-25) 
\ArrowLine(15,0)(30,25) 
\ArrowLine(30,-25)(15,0) 
\Photon(-15,0)(15,0){2}{6}
\end{picture}}
}
\hspace*{.8cm}
=
\mbox{IR fin.}
\]
\vspace*{1cm}
\[
\hspace*{0.44cm}
\vcenter{
\hbox{
  \begin{picture}(0,0)(0,0)
\SetScale{.8}
  \SetWidth{.5}
\ArrowLine(-30,25)(-15,0) 
\ArrowLine(-15,0)(-30,-25) 
\ArrowLine(15,0)(30,25) 
\ArrowLine(30,-25)(15,0) 
\Photon(-15,0)(15,0){2}{6}
\end{picture}}  
}
\hspace{0.8cm}
\times
\hspace{1.cm}
\vcenter{\hbox{
  \begin{picture}(0,0)(0,0)
\SetScale{.8}
  \SetWidth{.5}
\ArrowLine(-30,25)(-25,16.67) 
\ArrowLine(-25,16.67)(-15,0) 
\ArrowLine(-15,0)(-25,-16.67) 
\ArrowLine(-25,-16.67)(-30,-25) 
\ArrowLine(15,0)(30,25) 
\ArrowLine(30,-25)(15,0) 
\Photon(-15,0)(15,0){2}{6}
\PhotonArc(-20,0)(15,106,254){2}{7}
\end{picture}}}
\hspace*{1.cm}
+
\hspace*{3.2cm}
  \vcenter{
\hbox{\begin{picture}(0,0)(0,0)
\SetScale{.8}
  \SetWidth{.5}
\ArrowLine(-30,25)(-15,0) 
\ArrowLine(-15,0)(-30,-25) 
\ArrowLine(15,0)(30,25) 
\ArrowLine(30,-25)(15,0) 
\Photon(-15,0)(15,0){2}{6}
\Text(-20,-2)[cb]{{\small $2(J_{11} + J_{12})$}}
\end{picture}}
}
\hspace{.8cm}
\times
\hspace{.8cm}
  \vcenter{
\hbox{\begin{picture}(0,0)(0,0)
\SetScale{.8}
  \SetWidth{.5}
\ArrowLine(-30,25)(-15,0) 
\ArrowLine(-15,0)(-30,-25) 
\ArrowLine(15,0)(30,25) 
\ArrowLine(30,-25)(15,0) 
\Photon(-15,0)(15,0){2}{6}
\end{picture}}
}
\hspace*{.8cm}
=
\mbox{IR fin.}
\]

\vspace*{.8cm}
\caption[]{\it Cancellation of the IR divergencies in the one-loop vertex diagrams.}
\label{1lcanc2}
\end{figure}

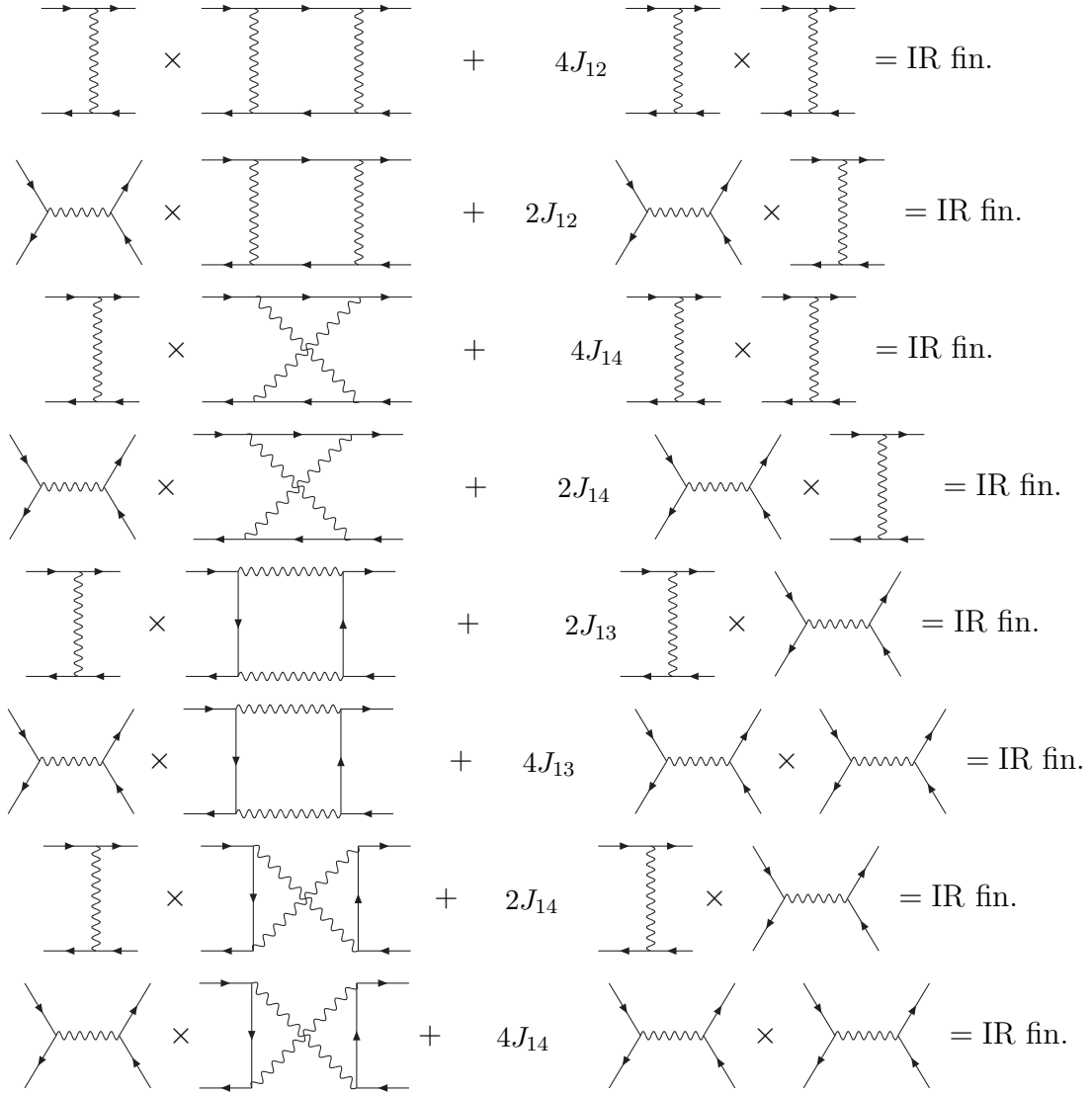
\begin{figure}
\vspace*{.6cm}
\[
\hspace*{1.cm}
\vcenter{
\hbox{
  \begin{picture}(0,0)(0,0)
\SetScale{.8}
  \SetWidth{.5}
\ArrowLine(-50,25)(-25,25)
\ArrowLine(-25,25)(-5,25)
\ArrowLine(-25,-25)(-50,-25)
\ArrowLine(-5,-25)(-25,-25)
\Photon(-25,25)(-25,-25){2}{10}
\end{picture}}  
}
\hspace{0.1cm}
\times
\hspace{1.4cm}
\vcenter{\hbox{
  \begin{picture}(0,0)(0,0)
\SetScale{.8}
  \SetWidth{.5}
\ArrowLine(-50,25)(-25,25)
\ArrowLine(-25,25)(25,25)
\ArrowLine(25,25)(50,25)
\Photon(-25,25)(-25,-25){2}{10}
\Photon(25,25)(25,-25){2}{10}
\ArrowLine(-25,-25)(-50,-25)
\ArrowLine(25,-25)(-25,-25)
\ArrowLine(50,-25)(25,-25)
\end{picture}}}
\hspace{2cm}
+
\hspace{3.2cm}
  \vcenter{
\hbox{\begin{picture}(0,0)(0,0)
\SetScale{.8}
  \SetWidth{.5}
\ArrowLine(-50,25)(-25,25)
\ArrowLine(-25,25)(-5,25)
\ArrowLine(-25,-25)(-50,-25)
\ArrowLine(-5,-25)(-25,-25)
\Photon(-25,25)(-25,-25){2}{10}
\Text(-20,-2)[cb]{{\small $4 J_{12}  $}}
\end{picture}}
}
\hspace{-0.1cm}
\times
\hspace{1.4cm}
  \vcenter{
\hbox{\begin{picture}(0,0)(0,0)
\SetScale{.8}
  \SetWidth{.5}
\ArrowLine(-50,25)(-25,25)
\ArrowLine(-25,25)(-5,25)
\ArrowLine(-25,-25)(-50,-25)
\ArrowLine(-5,-25)(-25,-25)
\Photon(-25,25)(-25,-25){2}{10}
\end{picture}}
}
=
\mbox{IR fin.}
\]
\vspace*{.6cm}
\[
\hspace*{0.5cm}
\vcenter{
\hbox{
  \begin{picture}(0,0)(0,0)
\SetScale{.8}
  \SetWidth{.5}
\ArrowLine(-30,25)(-15,0) 
\ArrowLine(-15,0)(-30,-25) 
\ArrowLine(15,0)(30,25) 
\ArrowLine(30,-25)(15,0) 
\Photon(-15,0)(15,0){2}{6}
\end{picture}}  
}
\hspace{1cm}
\times
\hspace{1.4cm}
\vcenter{\hbox{
  \begin{picture}(0,0)(0,0)
\SetScale{.8}
  \SetWidth{.5}
\ArrowLine(-50,25)(-25,25)
\ArrowLine(-25,25)(25,25)
\ArrowLine(25,25)(50,25)
\Photon(-25,25)(-25,-25){2}{10}
\Photon(25,25)(25,-25){2}{10}
\ArrowLine(-25,-25)(-50,-25)
\ArrowLine(25,-25)(-25,-25)
\ArrowLine(50,-25)(25,-25)
\end{picture}}}
\hspace{2cm}
+
\hspace{2.5cm}
  \vcenter{
\hbox{\begin{picture}(0,0)(0,0)
\SetScale{.8}
  \SetWidth{.5}
\ArrowLine(-30,25)(-15,0) 
\ArrowLine(-15,0)(-30,-25) 
\ArrowLine(15,0)(30,25) 
\ArrowLine(30,-25)(15,0) 
\Photon(-15,0)(15,0){2}{6}
\Text(-17,-2)[cb]{{\small $2 J_{12} $}}
\end{picture}}
}
\hspace{1cm}
\times
\hspace{1.4cm}
  \vcenter{
\hbox{\begin{picture}(0,0)(0,0)
\SetScale{.8}
  \SetWidth{.5}
\ArrowLine(-50,25)(-25,25)
\ArrowLine(-25,25)(-5,25)
\ArrowLine(-25,-25)(-50,-25)
\ArrowLine(-5,-25)(-25,-25)
\Photon(-25,25)(-25,-25){2}{10}
\end{picture}}
}
=
\mbox{IR fin.}
\]
\vspace*{.6cm}
\[
\hspace*{1.06cm}
\vcenter{
\hbox{
  \begin{picture}(0,0)(0,0)
\SetScale{.8}
  \SetWidth{.5}
\ArrowLine(-50,25)(-25,25)
\ArrowLine(-25,25)(-5,25)
\ArrowLine(-25,-25)(-50,-25)
\ArrowLine(-5,-25)(-25,-25)
\Photon(-25,25)(-25,-25){2}{10}
\end{picture}}  
}
\hspace{0.1cm}
\times
\hspace{1.5cm}
\vcenter{\hbox{\begin{picture}(0,0)(0,0)
\SetScale{.8}
  \SetWidth{.5}
\ArrowLine(-50,25)(-25,25)
\ArrowLine(-25,25)(25,25)
\ArrowLine(25,25)(50,25)
\Photon(-25,25)(25,-25){2}{10}
\Photon(25,25)(-25,-25){2}{10}
\ArrowLine(-25,-25)(-50,-25)
\ArrowLine(25,-25)(-25,-25)
\ArrowLine(50,-25)(25,-25)
\end{picture}}}
\hspace{2cm}
+
\hspace{3.2cm}
  \vcenter{
\hbox{\begin{picture}(0,0)(0,0)
\SetScale{.8}
  \SetWidth{.5}
\ArrowLine(-50,25)(-25,25)
\ArrowLine(-25,25)(-5,25)
\ArrowLine(-25,-25)(-50,-25)
\ArrowLine(-5,-25)(-25,-25)
\Photon(-25,25)(-25,-25){2}{10}
\Text(-18,-2)[cb]{{\small $4 J_{14} $}}
\end{picture}}
}
\hspace{-0.1cm}
\times
\hspace{1.4cm}
  \vcenter{
\hbox{\begin{picture}(0,0)(0,0)
\SetScale{.8}
  \SetWidth{.5}
\ArrowLine(-50,25)(-25,25)
\ArrowLine(-25,25)(-5,25)
\ArrowLine(-25,-25)(-50,-25)
\ArrowLine(-5,-25)(-25,-25)
\Photon(-25,25)(-25,-25){2}{10}
\end{picture}}
}
=
\mbox{IR fin.}
\]
\vspace*{.6cm}
\[
\hspace*{0.94cm}
\vcenter{
\hbox{
  \begin{picture}(0,0)(0,0)
\SetScale{.8}
  \SetWidth{.5}
\ArrowLine(-30,25)(-15,0) 
\ArrowLine(-15,0)(-30,-25) 
\ArrowLine(15,0)(30,25) 
\ArrowLine(30,-25)(15,0) 
\Photon(-15,0)(15,0){2}{6}
\end{picture}}  
}
\hspace{1cm}
\times
\hspace{1.52cm}
\vcenter{\hbox{\begin{picture}(0,0)(0,0)
\SetScale{.8}
  \SetWidth{.5}
\ArrowLine(-50,25)(-25,25)
\ArrowLine(-25,25)(25,25)
\ArrowLine(25,25)(50,25)
\Photon(-25,25)(25,-25){2}{10}
\Photon(25,25)(-25,-25){2}{10}
\ArrowLine(-25,-25)(-50,-25)
\ArrowLine(25,-25)(-25,-25)
\ArrowLine(50,-25)(25,-25)
\end{picture}
  }}
\hspace{2.cm}
+
\hspace{3cm}
  \vcenter{
\hbox{\begin{picture}(0,0)(0,0)
\SetScale{.8}
  \SetWidth{.5}
\ArrowLine(-30,25)(-15,0) 
\ArrowLine(-15,0)(-30,-25) 
\ArrowLine(15,0)(30,25) 
\ArrowLine(30,-25)(15,0) 
\Photon(-15,0)(15,0){2}{6}
\Text(-18,-2)[cb]{{\small $2  J_{14} $}}
\end{picture}}
}
\hspace{1cm}
\times
\hspace{1.4cm}
  \vcenter{
\hbox{\begin{picture}(0,0)(0,0)
\SetScale{.8}
  \SetWidth{.5}
\ArrowLine(-50,25)(-25,25)
\ArrowLine(-25,25)(-5,25)
\ArrowLine(-25,-25)(-50,-25)
\ArrowLine(-5,-25)(-25,-25)
\Photon(-25,25)(-25,-25){2}{10}
\end{picture}}
}
=
\mbox{IR fin.}
\]
\vspace*{.6cm}
\[
\hspace*{1.42cm}
\vcenter{
\hbox{
  \begin{picture}(0,0)(0,0)
\SetScale{.8}
  \SetWidth{.5}
\ArrowLine(-50,25)(-25,25)
\ArrowLine(-25,25)(-5,25)
\ArrowLine(-25,-25)(-50,-25)
\ArrowLine(-5,-25)(-25,-25)
\Photon(-25,25)(-25,-25){2}{10}
\end{picture}}  
}
\hspace{0.1cm}
\times
\hspace{1.4cm}
\vcenter{\hbox{
\begin{picture}(0,0)(0,0)
\SetScale{.8}
\SetWidth{.5}
\ArrowLine(-50,25)(-25,25)
\ArrowLine(25,25)(50,25)
\Photon(-25,-25)(25,-25){2}{10}
\Photon(-25,25)(25,25){2}{10}
\ArrowLine(-25,-25)(-50,-25)
\ArrowLine(50,-25)(25,-25)
\ArrowLine(-25,25)(-25,-25)
\ArrowLine(25,-25)(25,25)
\end{picture}
  }}
\hspace{2.cm}
+
\hspace{3.2cm}
  \vcenter{
\hbox{\begin{picture}(0,0)(0,0)
\SetScale{.8}
  \SetWidth{.5}
\ArrowLine(-50,25)(-25,25)
\ArrowLine(-25,25)(-5,25)
\ArrowLine(-25,-25)(-50,-25)
\ArrowLine(-5,-25)(-25,-25)
\Photon(-25,25)(-25,-25){2}{10}
\Text(-18,-2)[cb]{{\small $2 J_{13} $}}
\end{picture}}
}
\hspace{-0.1cm}
\times
\hspace{1.1cm}
  \vcenter{
\hbox{\begin{picture}(0,0)(0,0)
\SetScale{.8}
  \SetWidth{.5}
\ArrowLine(-30,25)(-15,0) 
\ArrowLine(-15,0)(-30,-25) 
\ArrowLine(15,0)(30,25) 
\ArrowLine(30,-25)(15,0) 
\Photon(-15,0)(15,0){2}{6}
\end{picture}}
}
\hspace*{1cm}
=
\mbox{IR fin.}
\]
%
\vspace*{.6cm}
\[
\hspace*{1.36cm}
\vcenter{
\hbox{\begin{picture}(0,0)(0,0)
\SetScale{.8}
  \SetWidth{.5}
\ArrowLine(-30,25)(-15,0) 
\ArrowLine(-15,0)(-30,-25) 
\ArrowLine(15,0)(30,25) 
\ArrowLine(30,-25)(15,0) 
\Photon(-15,0)(15,0){2}{6}
\end{picture}}
}
\hspace*{1.cm}
\times
\hspace*{1.4cm}
\vcenter{\hbox{
\begin{picture}(0,0)(0,0)
\SetScale{.8}
\SetWidth{.5}
\ArrowLine(-50,25)(-25,25)
\ArrowLine(25,25)(50,25)
\Photon(-25,-25)(25,-25){2}{10}
\Photon(-25,25)(25,25){2}{10}
\ArrowLine(-25,-25)(-50,-25)
\ArrowLine(50,-25)(25,-25)
\ArrowLine(-25,25)(-25,-25)
\ArrowLine(25,-25)(25,25)
\end{picture}
  }}
\hspace*{2.cm}
+
\hspace*{3cm}
  \vcenter{
\hbox{\begin{picture}(0,0)(0,0)
\SetScale{.8}
  \SetWidth{.5}
\ArrowLine(-30,25)(-15,0) 
\ArrowLine(-15,0)(-30,-25) 
\ArrowLine(15,0)(30,25) 
\ArrowLine(30,-25)(15,0) 
\Photon(-15,0)(15,0){2}{6}
\Text(-20,-2)[cb]{{\small $4 J_{13} $}}
\end{picture}}
}
\hspace{1cm}
\times
\hspace{1.1cm}
  \vcenter{
\hbox{\begin{picture}(0,0)(0,0)
\SetScale{.8}
  \SetWidth{.5}
\ArrowLine(-30,25)(-15,0) 
\ArrowLine(-15,0)(-30,-25) 
\ArrowLine(15,0)(30,25) 
\ArrowLine(30,-25)(15,0) 
\Photon(-15,0)(15,0){2}{6}
\end{picture}}
}
\hspace*{1cm}
=
\mbox{IR fin.}
\]
\vspace*{.6cm}
\[
\hspace*{1.36cm}
\vcenter{
\hbox{
  \begin{picture}(0,0)(0,0)
\SetScale{.8}
  \SetWidth{.5}
\ArrowLine(-50,25)(-25,25)
\ArrowLine(-25,25)(-5,25)
\ArrowLine(-25,-25)(-50,-25)
\ArrowLine(-5,-25)(-25,-25)
\Photon(-25,25)(-25,-25){2}{10}
\end{picture}}  
}
\hspace{0.1cm}
\times
\hspace{1.5cm}
\vcenter{\hbox{\begin{picture}(0,0)(0,0)
\SetScale{.8}
  \SetWidth{.5}
\ArrowLine(-50,25)(-25,25)
\ArrowLine(25,25)(50,25)
\Photon(-25,25)(25,-25){2}{10}
\Photon(-25,-25)(25,25){2}{10}
\ArrowLine(-25,-25)(-50,-25)
\ArrowLine(50,-25)(25,-25)
\ArrowLine(-25,25)(-25,-25)
\ArrowLine(25,-25)(25,25)
\end{picture}
  }}
\hspace{1.5cm}
+
\hspace{3.2cm}
  \vcenter{
\hbox{\begin{picture}(0,0)(0,0)
\SetScale{.8}
  \SetWidth{.5}
\ArrowLine(-50,25)(-25,25)
\ArrowLine(-25,25)(-5,25)
\ArrowLine(-25,-25)(-50,-25)
\ArrowLine(-5,-25)(-25,-25)
\Photon(-25,25)(-25,-25){2}{10}
\Text(-23,-2)[cb]{{\small $2 J_{14}$}}
\end{picture}}
}
\hspace{-0.1cm}
\times
\hspace{1.1cm}
  \vcenter{
\hbox{\begin{picture}(0,0)(0,0)
\SetScale{.8}
  \SetWidth{.5}
\ArrowLine(-30,25)(-15,0) 
\ArrowLine(-15,0)(-30,-25) 
\ArrowLine(15,0)(30,25) 
\ArrowLine(30,-25)(15,0) 
\Photon(-15,0)(15,0){2}{6}
\end{picture}}
}
\hspace*{1cm}
=
\mbox{IR fin.}
\]
%
\vspace*{.6cm}
\[
\hspace*{1.36cm}
\vcenter{
\hbox{\begin{picture}(0,0)(0,0)
\SetScale{.8}
  \SetWidth{.5}
\ArrowLine(-30,25)(-15,0) 
\ArrowLine(-15,0)(-30,-25) 
\ArrowLine(15,0)(30,25) 
\ArrowLine(30,-25)(15,0) 
\Photon(-15,0)(15,0){2}{6}
\end{picture}}
}
\hspace*{1.cm}
\times
\hspace*{1.4cm}
\vcenter{\hbox{\begin{picture}(0,0)(0,0)
\SetScale{.8}
  \SetWidth{.5}
\ArrowLine(-50,25)(-25,25)
\ArrowLine(25,25)(50,25)
\Photon(-25,25)(25,-25){2}{10}
\Photon(-25,-25)(25,25){2}{10}
\ArrowLine(-25,-25)(-50,-25)
\ArrowLine(50,-25)(25,-25)
\ArrowLine(-25,25)(-25,-25)
\ArrowLine(25,-25)(25,25)
\end{picture}
  }}
\hspace*{1.3cm}
+ 
\hspace*{3cm}
  \vcenter{
\hbox{\begin{picture}(0,0)(0,0)
\SetScale{.8}
  \SetWidth{.5}
\ArrowLine(-30,25)(-15,0) 
\ArrowLine(-15,0)(-30,-25) 
\ArrowLine(15,0)(30,25) 
\ArrowLine(30,-25)(15,0) 
\Photon(-15,0)(15,0){2}{6}
\Text(-20,-2)[cb]{{\small $4 J_{14} $}}
\end{picture}}
}
\hspace{1cm}
\times
\hspace{1.1cm}
  \vcenter{
\hbox{\begin{picture}(0,0)(0,0)
\SetScale{.8}
  \SetWidth{.5}
\ArrowLine(-30,25)(-15,0) 
\ArrowLine(-15,0)(-30,-25) 
\ArrowLine(15,0)(30,25) 
\ArrowLine(30,-25)(15,0) 
\Photon(-15,0)(15,0){2}{6}
\end{picture}}
}
\hspace*{1cm}
=
\mbox{IR fin.}
\]
\vspace*{.8cm}
\caption[]{\it Cancellation of the IR divergencies in the one-loop box diagrams.}
\label{1lcanc}
\end{figure}

At this stage, it is convenient to introduce the quantity 
\be
\frac{d \sigma_0^{D}(s,t,m^2)}{d \Omega} = 
\frac{d \sigma_0 (s,t,m^2)}{d \Omega}  + (D-4) 
\frac{d \sigma_0^{(D-4)} (s,t,m^2)}{d \Omega} +
{\mathcal O} \Bigl( (D-4)^2\Bigr) \, ,
\label{bornD1}
\ee
where:
\bea
\frac{d \sigma_0^{(D-4)} (s,t,m^2)}{d \Omega} &=&
\frac{\alpha}{s} \Biggl\{ \frac{1}{s^2}
 \left[ \frac{s^2}{4}\right] +
\frac{1}{t^2} \left[ \frac{t^2}{4}\right] \nn \\ 
& &+ \frac{1}{s t} \left[ \frac{1}{2} (s
+ t)^2 - 
\frac{1}{2} s t - m^2 (s + t) \right] 
\Biggr\} \, .
\label{bornD2}
\eea

$d \sigma_0^{D}/d \Omega$ is the Born Bhabha scattering cross-section obtained 
by calculating the traces over the Dirac indices in $D$ dimensions. The contribution 
of the $s$- and $t$- channel diagrams and of their interference to the r.h.s. of 
Eqs.~(\ref{bornD1},\ref{bornD2}) is explicit. It is then straightforward to show that, in the 
soft photon approximation, the contribution of the diagrams in Fig.~\ref{tlph} to 
the unpolarized differential cross-section is given by 
\be
\left( \frac{\alpha}{\pi} \right) 
\frac{d \sigma^{S}_1(s,t,m^2)}{d \Omega} = \left( \frac{\alpha}{\pi} \right)
\frac{d \sigma_0^{D}(s,t,m^2)}{d \Omega} \, \sum_{i,j=1}^4  J_{ij} \, , 
\label{a3recs}
\ee 
where the IR divergent quantity $J_{ij}$ is defined as
\be
J_{ij} = \epsilon_i \epsilon_j \, \left(p_i \cdot p_j \right) \, I_{ij}\, ,
\ee
with $\epsilon_i = +1$ for $i=1,4$ and $\epsilon_i = -1$ for $i=2,3$,
and where $I_{ij}$ indicates the integral:
\be
I_{ij} = \frac{1}{\Gamma\left(3-\frac{D}{2}\right) \pi^{(D-4)/2}}
 \frac{m^{D-4}}{4 \pi^2} \int^{\omega} \frac{d^D k}{k_0} \frac{1}{(p_i \cdot k) 
 (p_j \cdot k)} \, . 
\label{Iij}
\ee
In Eq.~(\ref{Iij}), $D$ is the dimensional regulator; furthermore the
superscript on the  integral sign indicates that the integration should be
taken over the region  $|\vec{k}| = k_0 <\omega$, with $\omega$ representing 
the cut-off on the unobserved soft-photon energy. The integral in
Eq.~(\ref{Iij}) can be evaluated according to the  standard technique discussed
in detail\footnote{The integral in Eq.~(\ref{Iij}) 
and the one in Ref.~\cite{GP} differ by a normalization factor. In particular, 
the integral with the normalization employed in this paper can be obtained by
multiplying Eq.~(1.230) in Ref.~\cite{GP} by the factor
\begin{displaymath}
1 - \frac{D-4}{2} \left( \gamma + \ln{\pi} + \ln{\left(\frac{m^2}{\mu^2} 
\right)} \right) + {\mathcal O}\left((D-4)^2\right)\,,
\end{displaymath}
where $\gamma$ is the Euler constant and $\mu$ the 't~Hooft scale.
Our choice of the normalization constants removes, in all the divergent
integrals, the finite terms associated with the use of dimensional regularization
and sets the 't~Hooft scale to $m$; this choice is consistent with the
normalization employed in Ref.~\cite{us2}.
}
 in Ref.~\cite{GP}. It is important to observe that the 
integrals $I_{ij}$ depend only on the scalar product $p_i \cdot p_j$ (aside
from an obvious dependence on $E$ and $m$), so that 
\bea
I_{ij} = I_{ji} \,, \quad I_{11} = I_{22}= I_{33} = I_{44} \, ,  \nn \\
I_{12} = I_{34} \,, \quad I_{13} = I_{24} \,, \quad I_{14} = I_{23} \, . 
\label{Iijsym}
\eea
Consequently, the quantities $J_{ij}$ also satisfy the same symmetry relations.
Therefore, Eq.~(\ref{a3recs}) becomes
\be
\left( \frac{\alpha}{\pi} \right)
\frac{d \sigma^{S}_1(s,t,m^2)}{d \Omega} = \left( \frac{\alpha}{\pi} \right)
\frac{d \sigma_0^D(s,t,m^2)}{d \Omega}  \, 4 \sum_{j=1}^{4} J_{1j} \, .
\label{aux1}
\ee
The explicit expressions of $I_{1j}$ ($j =1,4$) can be found in Appendix 
\ref{app1}.

The IR pole that originates from the integral $I_{1j}$ is multiplied, in
Eq.~(\ref{aux1}), by the terms proportional to $(D-4)$ in $\sigma_0^{D}$;
this product provides a finite contribution to $\sigma_1^S$.
Terms proportional to $(D-4)$ in $\sigma_1^S$ must be neglected.

It is useful to understand how the cancellation of the IR poles works from a
diagrammatic point of view. Fig.~\ref{1lcanc2} and Fig.~\ref{1lcanc}
schematically  describe the situation. The contribution to the differential
cross-section of the interference of the two diagrams shown in the first term
of each line is IR divergent. This IR divergence cancels against the
contribution to the real cross-section given by the second term in each line,
where the product of the two tree-level diagrams represents the contribution of
their interference to the cross-section in Born's approximation. We remind the
reader that the one-loop photon self-energy diagrams are IR finite.


\boldmath
\subsection{Real Corrections at Order $\alpha^4 (N_F = 1)$}
\unboldmath

The order $\alpha^4 (N_F = 1)$ contribution to the cross-section is given by:
\be
\left( \frac{\alpha}{\pi} \right)^2 \frac{d \sigma_2^{T}(s,t,m^2)}{d \Omega} = 
\left( \frac{\alpha}{\pi} \right)^2 
\left[ \frac{d \sigma^{V}_2(s,t,m^2)}{d \Omega}
+ \frac{d \sigma^{S}_2(s,t,m^2)}{d \Omega} \right]  \, ,
\label{2LOOP}
\ee
where the two-loop virtual cross-section $d \sigma^{V}_2/d \Omega$ can be found in
Eq.~(68) of \cite{us2}.

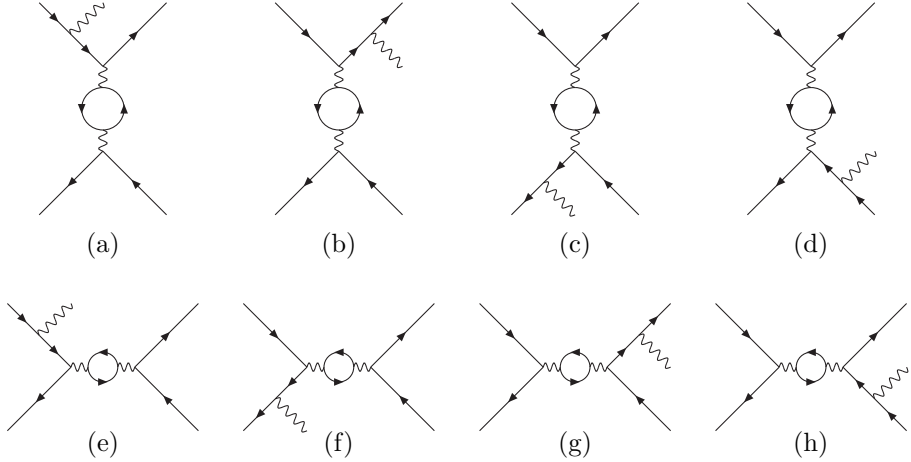
\begin{figure}
\vspace*{.3cm}
\[\vcenter{
\hbox{
  \begin{picture}(0,0)(0,0)
\SetScale{.8}
  \SetWidth{.5}
\ArrowLine(0,20)(30,50)
\ArrowLine(-30,50)(-15,35)
\ArrowLine(-15,35)(0,20)
\ArrowLine(30,-50)(0,-20)
\ArrowLine(0,-20)(-30,-50)
\Photon(0,-20)(0,-10){2}{2}
\Photon(0,10)(0,20){2}{2}
\ArrowArc(0,0)(10,-90,90)
\ArrowArc(0,0)(10,90,270)
\Photon(-15,35)(0,50){2}{4}
%
\Text(0,-20)[cb]{{\footnotesize (a)}}
\end{picture}}  
}
\hspace{3cm}
  \vcenter{
\hbox{
  \begin{picture}(0,0)(0,0)
\SetScale{.8}
  \SetWidth{.5}
\ArrowLine(0,20)(15,35)
\ArrowLine(15,35)(30,50)
\ArrowLine(-30,50)(0,20)
\ArrowLine(30,-50)(0,-20)
\ArrowLine(0,-20)(-30,-50)
\Photon(0,-20)(0,-10){2}{2}
\Photon(0,10)(0,20){2}{2}
\ArrowArc(0,0)(10,-90,90)
\ArrowArc(0,0)(10,90,270)
\Photon(15,35)(30,20){2}{4}
\Text(0,-20)[cb]{{\footnotesize (b)}}
\end{picture}}  
}
\hspace{3cm}
  \vcenter{
\hbox{
  \begin{picture}(0,0)(0,0)
\SetScale{.8}
  \SetWidth{.5}
\ArrowLine(0,20)(30,50)
\ArrowLine(-30,50)(0,20)
\ArrowLine(-15,-35)(-30,-50)
\ArrowLine(0,-20)(-15,-35)
\ArrowLine(30,-50)(0,-20)
\Photon(0,-20)(0,-10){2}{2}
\Photon(0,10)(0,20){2}{2}
\ArrowArc(0,0)(10,-90,90)
\ArrowArc(0,0)(10,90,270)
\Photon(-15,-35)(0,-50){2}{4}
\Text(0,-20)[cb]{{\footnotesize (c)}}
\end{picture}}  
}
\hspace{3cm}
  \vcenter{
\hbox{
  \begin{picture}(0,0)(0,0)
\SetScale{.8}
  \SetWidth{.5}
\ArrowLine(0,20)(30,50)
\ArrowLine(-30,50)(0,20)
\ArrowLine(0,-20)(-30,-50)
\ArrowLine(30,-50)(15,-35)
\ArrowLine(15,-35)(0,-20)
\Photon(0,-20)(0,-10){2}{2}
\Photon(0,10)(0,20){2}{2}
\ArrowArc(0,0)(10,-90,90)
\ArrowArc(0,0)(10,90,270)
\Photon(15,-35)(30,-20){2}{4}
\Text(0,-20)[cb]{{\footnotesize (d)}}
\end{picture}}  
}
\]
\vspace*{2cm}
\[\vcenter{
\hbox{
  \begin{picture}(0,0)(0,0)
\SetScale{.8}
  \SetWidth{.5}
\ArrowLine(-45,30)(-30,15)
\ArrowLine(-30,15)(-15,0)
\ArrowLine(-15,0)(-45,-30)
\ArrowLine(15,0)(45,30)
\ArrowLine(45,-30)(15,0)
\Photon(-15,0)(-7,0){2}{2}
\Photon(7,0)(15,0){2}{2}
\ArrowArc(0,0)(7,0,180)
\ArrowArc(0,0)(7,180,360)
\Photon(-30,15)(-15,30){2}{4}
\Text(0,-12)[cb]{{\footnotesize (e)}}
\end{picture}}  
}
\hspace{3cm}
  \vcenter{
\hbox{
  \begin{picture}(0,0)(0,0)
\SetScale{.8}
  \SetWidth{.5}
\ArrowLine(-45,30)(-15,0)
\ArrowLine(-15,0)(-30,-15)
\ArrowLine(-30,-15)(-45,-30)
\ArrowLine(15,0)(45,30)
\ArrowLine(45,-30)(15,0)
\Photon(-15,0)(-7,0){2}{2}
\Photon(7,0)(15,0){2}{2}
\ArrowArc(0,0)(7,0,180)
\ArrowArc(0,0)(7,180,360)
\Photon(-30,-15)(-15,-30){2}{4}
\Text(0,-12)[cb]{{\footnotesize (f)}}
\end{picture}}  
}
\hspace{3cm}
  \vcenter{
\hbox{
  \begin{picture}(0,0)(0,0)
\SetScale{.8}
  \SetWidth{.5}
\ArrowLine(-45,30)(-15,0)
\ArrowLine(-15,0)(-45,-30)
\ArrowLine(15,0)(30,15)
\ArrowLine(30,15)(45,30)
\ArrowLine(45,-30)(15,0)
\Photon(-15,0)(-7,0){2}{2}
\Photon(7,0)(15,0){2}{2}
\ArrowArc(0,0)(7,0,180)
\ArrowArc(0,0)(7,180,360)
\Photon(30,15)(45,0){2}{4}
\Text(0,-12)[cb]{{\footnotesize (g)}}
\end{picture}}  
}
\hspace{3cm}
  \vcenter{
\hbox{
  \begin{picture}(0,0)(0,0)
\SetScale{.8}
  \SetWidth{.5}
\ArrowLine(-45,30)(-15,0)
\ArrowLine(-15,0)(-45,-30)
\ArrowLine(15,0)(45,30)
\ArrowLine(45,-30)(30,-15)
\ArrowLine(30,-15)(15,0)
\Photon(-15,0)(-7,0){2}{2}
\Photon(7,0)(15,0){2}{2}
\ArrowArc(0,0)(7,0,180)
\ArrowArc(0,0)(7,180,360)
\Photon(30,-15)(45,0){2}{4}
\Text(0,-12)[cb]{{\footnotesize (h)}}
\end{picture}}  
}
\]
\vspace*{.6cm}
\caption[]{\it Diagrams contributing to the real corrections at order 
$\alpha^4(N_F = 1)$.}
\label{2lreal}
\end{figure}

The IR divergencies present in the Bhabha scattering differential cross-section
at order $\alpha^4 (N_F = 1)$ cancel against the contribution to the real 
cross-section of the interference of the diagrams in Fig.~\ref{2lreal} with the
single photon emission tree-level diagrams  in Fig.~\ref{tlph}.
%
In discussing the soft corrections to the cross-section at order 
$\alpha^4 (N_F = 1)$  it is convenient to introduce the quantity
\be
\frac{d \sigma_1^{D}(s,t,m^2)}{d \Omega}\Big|_{(1l,S)}  = 
\frac{d \sigma_1^{V}(s,t,m^2)}{d \Omega}\Big|_{(1l,S)} +
(D-4)\frac{d \sigma_1^{(D-4)}(s,t,m^2)}{d \Omega} \Big|_{(1l,S)} 
\label{newV} \, .
\ee
The first term in the r.h.s.  of the equation above is the
contribution to the virtual cross-section 
of the one-loop self-energy
diagrams and corresponding counter-term diagrams:
\bea
\frac{d \sigma_{1}^{V}(s,t,m^2)}{d \Omega} \Big|_{(1l,S)} &=&  
\frac{\alpha^2}{s} \Bigg\{
\frac{1}{s^2}\left[s t + \frac{s^2}{2} + (t - 2 m^2)^2\right] 
2 \mbox{Re}\Pi^{(1l,0)}_0(s)  \nn\\  
& & + \frac{1}{t^2}\left[s t + \frac{t^2}{2} + (s - 2 m^2)^2\right]  
2 \Pi^{(1l,0)}_0(t)  \nn\\ 
& & + \frac{1}{s t}\left[(s + t)^2 - 4 m^4\right]
 \left(\mbox{Re}\Pi^{(1l,0)}_0(s) + 
  \Pi^{(1l,0)}_0(t)
 \right) 
\Bigg\} ,
\eea
where $\Pi^{(1l,0)}_0$  is the UV renormalized photon self-energy.
 The latter quantity has been 
discussed in detail in Section~4 of \cite{us2}. 
The second term in the r.h.s. of Eq.~(\ref{newV}) is: 
\bea
\frac{d \sigma_{1}^{(D-4)}(s,t,m^2)}{d \Omega} \Big|_{(1l,S)}\! \! \! \! &=&  \! \! 
\frac{\alpha^2}{s} \Bigg\{
\frac{1}{s^2}\left[s t + \frac{s^2}{2} + (t - 2 m^2)^2\right] 
2 \mbox{Re}\Pi^{(1l,1)}_0(s)  \nn\\  
\! \!& & \! \!+ \frac{1}{t^2}\left[s t + \frac{t^2}{2} + (s - 2 m^2)^2\right]  
2 \Pi^{(1l,1)}_0(t)  \nn\\ 
\! \!& & \! \!+ \frac{1}{s t}\left[(s + t)^2 - 4 m^4\right]
 \left(\mbox{Re}\Pi^{(1l,1)}_0(s) + 
  \Pi^{(1l,1)}_0(t)\right)  \nn \\
\! \!& & \! \!+\frac{\mbox{Re}\Pi^{(1l,0)}_0(s) + \Pi^{(1l,0)}_0(t)}{2} + 
\frac{1}{2} \left[ (s +t)^2 \! - \! s t  \! - \! 2 m^2 (s\! +\! t)
\right] \times \nn \\
\! \!& &\! \!\times\left(\mbox{Re}\Pi^{(1l,0)}_0(s) + 
  \Pi^{(1l,0)}_0(t)\right)
\Bigg\} \, , 
\label{amp1lS}
\eea
where $\Pi^{(1l,1)}_0$  is the term proportional to $(D-4)$
in the expantion of the  renormalized photon self-energy (the explicit
expression of this quantity can be found in the appendix of  \cite{us2}).

The contribution of the soft photon emission to the real corrections to the 
Bhabha scattering cross-section at order $\alpha^4 (N_F = 1)$ is given by
\be
\left(\frac{\alpha}{\pi}\right)^2
\frac{d \sigma^{S}_2(s,t,m^2)}{d \Omega} = \left(\frac{\alpha}{\pi}\right)^2 
\left(\frac{d \sigma_1^{D}(s,t,m^2)}{d \Omega} \Big|_{(1l,S)}  
\right) \, 4 \sum_{j=1}^{4} J_{1j}
\, ,
\label{above2}
\ee
\begin{figure}
\vspace*{.3cm}
\[
\hspace*{8mm}
\vcenter{
\hbox{
  \begin{picture}(0,0)(0,0)
\SetScale{.8}
  \SetWidth{.5}
\ArrowLine(-25,25)(-15,21) 
\ArrowLine(-15,21)(0,15) 
\ArrowLine(15,21)(25,25) 
\ArrowLine(0,15)(15,21) 
\ArrowLine(0,-15)(-25,-25)
\ArrowLine(25,-25)(0,-15)
\Photon(0,15)(0,10){2}{1.5}
\Photon(0,-10)(0,-15){2}{1.5}
\ArrowArc(0,0)(10,-90,90)
\ArrowArc(0,0)(10,90,270)
\PhotonArc(0,15)(15,22.5,157.5){2}{6}

\end{picture}}  
}
\hspace{.7cm}
\times
\hspace{1.4cm}
  \vcenter{
\hbox{\begin{picture}(0,0)(0,0)
\SetScale{.8}
  \SetWidth{.5}
\ArrowLine(-50,25)(-25,25)
\ArrowLine(-25,25)(-5,25)
\ArrowLine(-25,-25)(-50,-25)
\ArrowLine(-5,-25)(-25,-25)
\Photon(-25,25)(-25,-25){2}{10}
\end{picture}}
}
%
\hspace{1cm}
+ 
\hspace{4.cm}
%
  \vcenter{
\hbox{
  \begin{picture}(0,0)(0,0)
\SetScale{.8}
  \SetWidth{.5}
\ArrowLine(-50,25)(-25,25)
\ArrowLine(-25,25)(-5,25)
\ArrowLine(-25,-25)(-50,-25)
\ArrowLine(-5,-25)(-25,-25)
\Photon(-25,25)(-25,10){2}{3}
\ArrowArc(-25,0)(10,-90,90)
\ArrowArc(-25,0)(10,90,270)
\Photon(-25,-10)(-25,-25){2}{3}
\Text(-24,-2)[cb]{{\small 2 $\left(J_{13} + J_{11} \right)$}}
\end{picture}}  
}
\hspace{-0.1cm}
\times
\hspace{1.4cm}
  \vcenter{
\hbox{\begin{picture}(0,0)(0,0)
\SetScale{.8}
  \SetWidth{.5}
\ArrowLine(-50,25)(-25,25)
\ArrowLine(-25,25)(-5,25)
\ArrowLine(-25,-25)(-50,-25)
\ArrowLine(-5,-25)(-25,-25)
\Photon(-25,25)(-25,-25){2}{10}
\end{picture}}
}
\hspace*{0.5cm}
=
\mbox{IR fin.}
\]
\vspace*{1cm}
\[
\hspace*{8mm}
\vcenter{
\hbox{
  \begin{picture}(0,0)(0,0)
\SetScale{.8}
  \SetWidth{.5}
\ArrowLine(-25,25)(-15,21) 
\ArrowLine(-15,21)(0,15) 
\ArrowLine(15,21)(25,25) 
\ArrowLine(0,15)(15,21) 
\ArrowLine(0,-15)(-25,-25)
\ArrowLine(25,-25)(0,-15)
\Photon(0,15)(0,10){2}{1.5}
\Photon(0,-10)(0,-15){2}{1.5}
\ArrowArc(0,0)(10,-90,90)
\ArrowArc(0,0)(10,90,270)
\PhotonArc(0,15)(15,22.5,157.5){2}{6}

%
\end{picture}}  
}
\hspace{0.7cm}
\times
\hspace{.8cm}
  \vcenter{
\hbox{
  \begin{picture}(0,0)(0,0)
\SetScale{.8}
  \SetWidth{.5}
\ArrowLine(-30,25)(-15,0) 
\ArrowLine(-15,0)(-30,-25) 
\ArrowLine(15,0)(30,25) 
\ArrowLine(30,-25)(15,0) 
\Photon(-15,0)(15,0){2}{6}
\end{picture}}  
}
%
\hspace{1.8cm}
+
\hspace{4.cm}
%
  \vcenter{
\hbox{\begin{picture}(0,0)(0,0)
\SetScale{.8}
  \SetWidth{.5}
\ArrowLine(-50,25)(-25,25)
\ArrowLine(-25,25)(-5,25)
\ArrowLine(-25,-25)(-50,-25)
\ArrowLine(-5,-25)(-25,-25)
\Photon(-25,25)(-25,10){2}{3}
\ArrowArc(-25,0)(10,-90,90)
\ArrowArc(-25,0)(10,90,270)
\Photon(-25,-10)(-25,-25){2}{3}
\Text(-26,-2)[cb]{{\small $\left(J_{13} +J_{11} \right)$}}
\end{picture}}
}
\hspace{-0.1cm}
\times
\hspace{1.1cm}
  \vcenter{
\hbox{\begin{picture}(0,0)(0,0)
\SetScale{.8}
  \SetWidth{.5}
\ArrowLine(-30,25)(-15,0) 
\ArrowLine(-15,0)(-30,-25) 
\ArrowLine(15,0)(30,25) 
\ArrowLine(30,-25)(15,0) 
\Photon(-15,0)(15,0){2}{6}
\end{picture}}
}
\hspace*{1cm}
=
\mbox{IR fin.}
\]
%
\vspace*{1cm}
\[
\hspace*{.6cm}
\vcenter{
\hbox{
  \begin{picture}(0,0)(0,0)
\SetScale{.8}
  \SetWidth{.5}
\ArrowLine(-30,25)(-25,16.67) 
\ArrowLine(-25,16.67)(-15,0) 
\ArrowLine(-15,0)(-25,-16.67) 
\ArrowLine(-25,-16.67)(-30,-25) 
\ArrowLine(15,0)(30,25) 
\ArrowLine(30,-25)(15,0)
 \ArrowArc(0,0)(10,0,180)
\ArrowArc(0,0)(10,180,360)
\Photon(-15,0)(-10,0){2}{1.5}
\Photon(10,0)(15,0){2}{1.5} 
\PhotonArc(-20,0)(15,106,254){2}{7}
%
\end{picture}}  
}
\hspace{1.cm}
\times
\hspace{1.2cm}
  \vcenter{
\hbox{
  \begin{picture}(0,0)(0,0)
\SetScale{.8}
  \SetWidth{.5}
\ArrowLine(-50,25)(-25,25)
\ArrowLine(-25,25)(-5,25)
\ArrowLine(-25,-25)(-50,-25)
\ArrowLine(-5,-25)(-25,-25)
\Photon(-25,25)(-25,-25){2}{10}
\end{picture}}  
}
%
\hspace{-.1cm}
+ 
\hspace{4.cm}
%
  \vcenter{
\hbox{\begin{picture}(0,0)(0,0)
\SetScale{.8}
  \SetWidth{.5}
\ArrowLine(-50,25)(-25,25)
\ArrowLine(-25,25)(-5,25)
\ArrowLine(-25,-25)(-50,-25)
\ArrowLine(-5,-25)(-25,-25)
\Photon(-25,25)(-25,-25){2}{10}
\Text(-24,-2)[cb]{{\small $\left(J_{12} + J_{11} \right)$}}
\end{picture}}
}
\hspace{-0.1cm}
\times
\hspace{0.9cm}
  \vcenter{
\hbox{
  \begin{picture}(0,0)(0,0)
\SetScale{.8}
  \SetWidth{.5}
\ArrowLine(-30,25)(-15,0) 
\ArrowLine(-15,0)(-30,-25) 
\ArrowLine(15,0)(30,25) 
\ArrowLine(30,-25)(15,0) 
\Photon(-15,0)(-10,0){2}{1.5}
\Photon(15,0)(10,0){2}{1.5}
\ArrowArc(0,0)(10,0,180)
\ArrowArc(0,0)(10,180,360)
\end{picture}}  
}
\hspace*{1cm}
=
\mbox{IR pole} \propto \zeta(2)
\]
\vspace*{1cm}
\[
\hspace*{9mm}
\vcenter{
\hbox{
  \begin{picture}(0,0)(0,0)
\SetScale{.8}
  \SetWidth{.5}
\ArrowLine(-30,25)(-25,16.67) 
\ArrowLine(-25,16.67)(-15,0) 
\ArrowLine(-15,0)(-25,-16.67) 
\ArrowLine(-25,-16.67)(-30,-25) 
\ArrowLine(15,0)(30,25) 
\ArrowLine(30,-25)(15,0)
 \ArrowArc(0,0)(10,0,180)
\ArrowArc(0,0)(10,180,360)
\Photon(-15,0)(-10,0){2}{1.5}
\Photon(10,0)(15,0){2}{1.5} 
\PhotonArc(-20,0)(15,106,254){2}{7}
%
\end{picture}}  
}
\hspace*{1.cm}
\times
\hspace*{0.8cm}
  \vcenter{
\hbox{
  \begin{picture}(0,0)(0,0)
\SetScale{.8}
  \SetWidth{.5}
\ArrowLine(-30,25)(-15,0) 
\ArrowLine(-15,0)(-30,-25) 
\ArrowLine(15,0)(30,25) 
\ArrowLine(30,-25)(15,0) 
\Photon(-15,0)(15,0){2}{6}
\end{picture}}  
}
%
\hspace*{1.cm}
+ 
\hspace*{3.6cm}
%
  \vcenter{
\hbox{\begin{picture}(0,0)(0,0)
\SetScale{.8}
  \SetWidth{.5}
\ArrowLine(-30,25)(-15,0) 
\ArrowLine(-15,0)(-30,-25) 
\ArrowLine(15,0)(30,25) 
\ArrowLine(30,-25)(15,0) 
\Photon(-15,0)(15,0){2}{6}
\Text(-22,-2)[cb]{{\small 2 $\left(J_{12} + J_{11} \right)$}}
\end{picture}}
}
\hspace*{.8cm}
\times
\hspace*{0.8cm}
  \vcenter{
\hbox{
  \begin{picture}(0,0)(0,0)
\SetScale{.8}
  \SetWidth{.5}
\ArrowLine(-30,25)(-15,0) 
\ArrowLine(-15,0)(-30,-25) 
\ArrowLine(15,0)(30,25) 
\ArrowLine(30,-25)(15,0) 
\Photon(-15,0)(-10,0){2}{1.5}
\Photon(15,0)(10,0){2}{1.5}
\ArrowArc(0,0)(10,0,180)
\ArrowArc(0,0)(10,180,360)
\end{picture}}  
}
\hspace*{1cm}
=
\mbox{IR pole} \propto \zeta(2)
\]
\vspace*{.6cm}
\caption[]{\it Cancellation of the IR divergencies of the two-loop $N_F=1$
reducible diagrams.}
\label{2lreducible}
\end{figure}
%
%
%
%
%
%
\begin{figure}
\vspace*{.3cm}
\[\vcenter{
\hbox{
  \begin{picture}(0,0)(0,0)
\SetScale{.8}
  \SetWidth{.5}
\ArrowLine(-25,25)(-15,21) 
\ArrowLine(-15,21)(0,15) 
\ArrowLine(15,21)(25,25) 
\ArrowLine(0,15)(15,21) 
\ArrowLine(0,-15)(-25,-25)
\ArrowLine(25,-25)(0,-15)
\Photon(0,15)(0,-15){2}{5}
\PhotonArc(0,15)(15,22.5,157.5){2}{6}
%
\end{picture}}  
}
\hspace*{.7cm}
\times
\hspace*{1.4cm}
  \vcenter{
\hbox{
  \begin{picture}(0,0)(0,0)
\SetScale{.8}
  \SetWidth{.5}
\ArrowLine(-50,25)(-25,25)
\ArrowLine(-25,25)(-5,25)
\ArrowLine(-25,-25)(-50,-25)
\ArrowLine(-5,-25)(-25,-25)
\Photon(-25,25)(-25,10){2}{3}
\ArrowArc(-25,0)(10,-90,90)
\ArrowArc(-25,0)(10,90,270)
\Photon(-25,-10)(-25,-25){2}{3}
\end{picture}}  
}
%
\hspace*{0.4cm}
+
\hspace*{4.cm}
%
  \vcenter{
\hbox{\begin{picture}(0,0)(0,0)
\SetScale{.8}
  \SetWidth{.5}
\ArrowLine(-50,25)(-25,25)
\ArrowLine(-25,25)(-5,25)
\ArrowLine(-25,-25)(-50,-25)
\ArrowLine(-5,-25)(-25,-25)
\Photon(-25,25)(-25,10){2}{3}
\ArrowArc(-25,0)(10,-90,90)
\ArrowArc(-25,0)(10,90,270)
\Photon(-25,-10)(-25,-25){2}{3}
\Text(-26,-2)[cb]{{\small $\left(J_{13} + J_{11} \right)$}}
\end{picture}}
}
\hspace*{0.1cm}
\times
\hspace*{1.4cm}
  \vcenter{
\hbox{\begin{picture}(0,0)(0,0)
\SetScale{.8}
  \SetWidth{.5}
\ArrowLine(-50,25)(-25,25)
\ArrowLine(-25,25)(-5,25)
\ArrowLine(-25,-25)(-50,-25)
\ArrowLine(-5,-25)(-25,-25)
\Photon(-25,25)(-25,-25){2}{10}
\end{picture}}
}
\hspace*{0.2cm}
=
\mbox{IR fin.}
\]
\vspace*{1cm}
\[\vcenter{
\hbox{
  \begin{picture}(0,0)(0,0)
\SetScale{.8}
  \SetWidth{.5}
\ArrowLine(-25,25)(-15,21) 
\ArrowLine(-15,21)(0,15) 
\ArrowLine(15,21)(25,25) 
\ArrowLine(0,15)(15,21) 
\ArrowLine(0,-15)(-25,-25)
\ArrowLine(25,-25)(0,-15)
\Photon(0,15)(0,-15){2}{5}
\PhotonArc(0,15)(15,22.5,157.5){2}{6}
%
\end{picture}}  
}
\hspace{.7cm}
\times
\hspace{.9cm}
  \vcenter{
\hbox{
  \begin{picture}(0,0)(0,0)
\SetScale{.8}
  \SetWidth{.5}
\ArrowLine(-30,25)(-15,0) 
\ArrowLine(-15,0)(-30,-25) 
\ArrowLine(15,0)(30,25) 
\ArrowLine(30,-25)(15,0) 
\Photon(-15,0)(-10,0){2}{1.5}
\Photon(15,0)(10,0){2}{1.5}
\ArrowArc(0,0)(10,0,180)
\ArrowArc(0,0)(10,180,360)
\end{picture}}  
}
%
\hspace{1.2cm}
+ 
\hspace{4.cm}
%
  \vcenter{
\hbox{\begin{picture}(0,0)(0,0)
\SetScale{.8}
  \SetWidth{.5}
\ArrowLine(-50,25)(-25,25)
\ArrowLine(-25,25)(-5,25)
\ArrowLine(-25,-25)(-50,-25)
\ArrowLine(-5,-25)(-25,-25)
\Photon(-25,25)(-25,-25){2}{10}
\Text(-26,-2)[cb]{{\small $\left(J_{13} + J_{11} \right)$}}
\end{picture}}
}
\hspace{-0.1cm}
\times
\hspace{1.1cm}
  \vcenter{
\hbox{\begin{picture}(0,0)(0,0)
\SetScale{.8}
  \SetWidth{.5}
\ArrowLine(-30,25)(-15,0) 
\ArrowLine(-15,0)(-30,-25) 
\ArrowLine(15,0)(30,25) 
\ArrowLine(30,-25)(15,0) 
\Photon(-15,0)(-10,0){2}{1.5}
\Photon(15,0)(10,0){2}{1.5}
\ArrowArc(0,0)(10,0,180)
\ArrowArc(0,0)(10,180,360)
\end{picture}}
}
\hspace*{1cm}
=
\mbox{IR fin.}
\]
%
\vspace*{1cm}
\[\vcenter{
\hbox{
  \begin{picture}(0,0)(0,0)
\SetScale{.8}
  \SetWidth{.5}

\ArrowLine(-30,25)(-25,16.67) 
\ArrowLine(-25,16.67)(-15,0) 
\ArrowLine(-15,0)(-25,-16.67) 
\ArrowLine(-25,-16.67)(-30,-25) 
\ArrowLine(15,0)(30,25) 
\ArrowLine(30,-25)(15,0) 
\Photon(-15,0)(15,0){2}{6}
\PhotonArc(-20,0)(15,106,254){2}{7}

%
\end{picture}}  
}
\hspace{1.cm}
\times
\hspace{1.2cm}
  \vcenter{
\hbox{
  \begin{picture}(0,0)(0,0)
\SetScale{.8}
  \SetWidth{.5}
\ArrowLine(-50,25)(-25,25)
\ArrowLine(-25,25)(-5,25)
\ArrowLine(-25,-25)(-50,-25)
\ArrowLine(-5,-25)(-25,-25)
\Photon(-25,25)(-25,10){2}{3}
\ArrowArc(-25,0)(10,-90,90)
\ArrowArc(-25,0)(10,90,270)
\Photon(-25,-10)(-25,-25){2}{3}
\end{picture}}  
}
%
\hspace{0.1cm}
+
\hspace{4.cm}
%
  \vcenter{
\hbox{\begin{picture}(0,0)(0,0)
\SetScale{.8}
  \SetWidth{.5}
\ArrowLine(-50,25)(-25,25)
\ArrowLine(-25,25)(-5,25)
\ArrowLine(-25,-25)(-50,-25)
\ArrowLine(-5,-25)(-25,-25)
\Photon(-25,25)(-25,10){2}{3}
\ArrowArc(-25,0)(10,-90,90)
\ArrowArc(-25,0)(10,90,270)
\Photon(-25,-10)(-25,-25){2}{3}
\Text(-26,-2)[cb]{{\small $\left(J_{12} + J_{11} \right)$}}
\end{picture}}
}
\hspace{-0.1cm}
\times
\hspace{0.9cm}
  \vcenter{
\hbox{\begin{picture}(0,0)(0,0)
\SetScale{.8}
  \SetWidth{.5}
\ArrowLine(-30,25)(-15,0) 
\ArrowLine(-15,0)(-30,-25) 
\ArrowLine(15,0)(30,25) 
\ArrowLine(30,-25)(15,0) 
\Photon(-15,0)(15,0){2}{6}
\end{picture}}
}
\hspace*{1cm}
=
\mbox{IR fin.} 
\]
\vspace*{1cm}
\[
\hspace*{1.1cm}
\vcenter{
\hbox{
  \begin{picture}(0,0)(0,0)
\SetScale{.8}
  \SetWidth{.5}
\ArrowLine(-30,25)(-25,16.67) 
\ArrowLine(-25,16.67)(-15,0) 
\ArrowLine(-15,0)(-25,-16.67) 
\ArrowLine(-25,-16.67)(-30,-25) 
\ArrowLine(15,0)(30,25) 
\ArrowLine(30,-25)(15,0) 
\Photon(-15,0)(15,0){2}{6}
\PhotonArc(-20,0)(15,106,254){2}{7}
%
\end{picture}}  
}
\hspace*{.8cm}
\times
\hspace*{.8cm}
  \vcenter{
\hbox{
  \begin{picture}(0,0)(0,0)
\SetScale{.8}
  \SetWidth{.5}
\ArrowLine(-30,25)(-15,0) 
\ArrowLine(-15,0)(-30,-25) 
\ArrowLine(15,0)(30,25) 
\ArrowLine(30,-25)(15,0) 
\Photon(-15,0)(-10,0){2}{1.5}
\Photon(15,0)(10,0){2}{1.5}
\ArrowArc(0,0)(10,0,180)
\ArrowArc(0,0)(10,180,360)
\end{picture}}  
}
%
\hspace*{1.2cm}
+
\hspace*{3.6cm}
%
  \vcenter{
\hbox{\begin{picture}(0,0)(0,0)
\SetScale{.8}
  \SetWidth{.5}
\ArrowLine(-30,25)(-15,0) 
\ArrowLine(-15,0)(-30,-25) 
\ArrowLine(15,0)(30,25) 
\ArrowLine(30,-25)(15,0) 
\Photon(-15,0)(-10,0){2}{1.5}
\Photon(15,0)(10,0){2}{1.5}
\ArrowArc(0,0)(10,0,180)
\ArrowArc(0,0)(10,180,360)
\Text(-22,-2)[cb]{{\small $\left(J_{12} + J_{11} \right)$}}
\end{picture}}
}
\hspace*{.8cm}
\times
\hspace*{0.8cm}
  \vcenter{
\hbox{\begin{picture}(0,0)(0,0)
\SetScale{.8}
  \SetWidth{.5}
\ArrowLine(-30,25)(-15,0) 
\ArrowLine(-15,0)(-30,-25) 
\ArrowLine(15,0)(30,25) 
\ArrowLine(30,-25)(15,0) 
\Photon(-15,0)(15,0){2}{6}
\end{picture}}
}
\hspace*{1cm}
=
\mbox{IR pole} \propto \zeta(2)
\]
\vspace*{.6cm}
\caption[]{\it Cancellation of the IR divergencies of the products of one-loop
self-energy and vertex diagrams.}
\label{onelVtimesonelS}
\end{figure}
%
\begin{figure}
\vspace*{.6cm}
\[\vcenter{
\hbox{
  \begin{picture}(0,0)(0,0)
\SetScale{.8}
  \SetWidth{.5}
\ArrowLine(-50,25)(-25,25)
\ArrowLine(-25,25)(-5,25)
\ArrowLine(-25,-25)(-50,-25)
\ArrowLine(-5,-25)(-25,-25)
\Photon(-25,25)(-25,10){2}{3}
\ArrowArc(-25,0)(10,-90,90)
\ArrowArc(-25,0)(10,90,270)
\Photon(-25,-10)(-25,-25){2}{3}
\end{picture}}  
}
\hspace{0.1cm}
\times
\hspace{1.4cm}
\vcenter{\hbox{
  \begin{picture}(0,0)(0,0)
\SetScale{.8}
  \SetWidth{.5}
\ArrowLine(-50,25)(-25,25)
\ArrowLine(-25,25)(25,25)
\ArrowLine(25,25)(50,25)
\Photon(-25,25)(-25,-25){2}{10}
\Photon(25,25)(25,-25){2}{10}
\ArrowLine(-25,-25)(-50,-25)
\ArrowLine(25,-25)(-25,-25)
\ArrowLine(50,-25)(25,-25)
\end{picture}}}
\hspace{2cm}
+
\hspace{3.2cm}
  \vcenter{
\hbox{\begin{picture}(0,0)(0,0)
\SetScale{.8}
  \SetWidth{.5}
\ArrowLine(-50,25)(-25,25)
\ArrowLine(-25,25)(-5,25)
\ArrowLine(-25,-25)(-50,-25)
\ArrowLine(-5,-25)(-25,-25)
\Photon(-25,25)(-25,10){2}{3}
\ArrowArc(-25,0)(10,-90,90)
\ArrowArc(-25,0)(10,90,270)
\Photon(-25,-10)(-25,-25){2}{3}  
\Text(-22,-2)[cb]{{\small $2 J_{12}  $}}
\end{picture}}
}
\hspace{-0.1cm}
\times
\hspace{1.4cm}
  \vcenter{
\hbox{\begin{picture}(0,0)(0,0)
\SetScale{.8}
  \SetWidth{.5}
\ArrowLine(-50,25)(-25,25)
\ArrowLine(-25,25)(-5,25)
\ArrowLine(-25,-25)(-50,-25)
\ArrowLine(-5,-25)(-25,-25)
\Photon(-25,25)(-25,-25){2}{10}
\end{picture}}
}
=
\mbox{IR fin.}
\]
\vspace*{.6cm}
\[\vcenter{
\hbox{
  \begin{picture}(0,0)(0,0)
\SetScale{.8}
  \SetWidth{.5}
\ArrowLine(-30,25)(-15,0) 
\ArrowLine(-15,0)(-30,-25) 
\ArrowLine(15,0)(30,25) 
\ArrowLine(30,-25)(15,0) 
\Photon(-15,0)(-10,0){2}{1.5}
\Photon(15,0)(10,0){2}{1.5}
\ArrowArc(0,0)(10,0,180)
\ArrowArc(0,0)(10,180,360)  
\end{picture}}  
}
\hspace{1cm}
\times
\hspace{1.4cm}
\vcenter{\hbox{
  \begin{picture}(0,0)(0,0)
\SetScale{.8}
  \SetWidth{.5}
\ArrowLine(-50,25)(-25,25)
\ArrowLine(-25,25)(25,25)
\ArrowLine(25,25)(50,25)
\Photon(-25,25)(-25,-25){2}{10}
\Photon(25,25)(25,-25){2}{10}
\ArrowLine(-25,-25)(-50,-25)
\ArrowLine(25,-25)(-25,-25)
\ArrowLine(50,-25)(25,-25)
\end{picture}}}
\hspace*{1cm}
+
\hspace*{2.5cm}
  \vcenter{
\hbox{\begin{picture}(0,0)(0,0)
\SetScale{.8}
  \SetWidth{.5}
\ArrowLine(-50,25)(-25,25)
\ArrowLine(-25,25)(-5,25)
\ArrowLine(-25,-25)(-50,-25)
\ArrowLine(-5,-25)(-25,-25)
\Photon(-25,25)(-25,-25){2}{10}
  \Text(-17,-2)[cb]{{\small $2 J_{12} $}}
\end{picture}}
}
\hspace*{.1cm}
\times
\hspace*{1.cm}
  \vcenter{
\hbox{\begin{picture}(0,0)(0,0)
\SetScale{.8}
  \SetWidth{.5}
 \ArrowLine(-30,25)(-15,0) 
\ArrowLine(-15,0)(-30,-25) 
\ArrowLine(15,0)(30,25) 
\ArrowLine(30,-25)(15,0) 
\Photon(-15,0)(-10,0){2}{1.5}
\Photon(15,0)(10,0){2}{1.5}
\ArrowArc(0,0)(10,0,180)
\ArrowArc(0,0)(10,180,360)
\end{picture}}
}
\hspace*{1.cm}
=
\mbox{IR Pole} \propto \zeta(2)
\]
\vspace*{.6cm}
\[\vcenter{
\hbox{
  \begin{picture}(0,0)(0,0)
\SetScale{.8}
  \SetWidth{.5}
\ArrowLine(-50,25)(-25,25)
\ArrowLine(-25,25)(-5,25)
\ArrowLine(-25,-25)(-50,-25)
\ArrowLine(-5,-25)(-25,-25)
\Photon(-25,25)(-25,10){2}{3}
\ArrowArc(-25,0)(10,-90,90)
\ArrowArc(-25,0)(10,90,270)
\Photon(-25,-10)(-25,-25){2}{3}
\end{picture}}  
}
\hspace{0.1cm}
\times
\hspace{1.4cm}
\vcenter{\hbox{\begin{picture}(0,0)(0,0)
\SetScale{.8}
  \SetWidth{.5}
\ArrowLine(-50,25)(-25,25)
\ArrowLine(-25,25)(25,25)
\ArrowLine(25,25)(50,25)
\Photon(-25,25)(25,-25){2}{10}
\Photon(25,25)(-25,-25){2}{10}
\ArrowLine(-25,-25)(-50,-25)
\ArrowLine(25,-25)(-25,-25)
\ArrowLine(50,-25)(25,-25)
\end{picture}}}
\hspace{2cm}
+
\hspace{3.2cm}
  \vcenter{
\hbox{\begin{picture}(0,0)(0,0)
\SetScale{.8}
  \SetWidth{.5}
 \ArrowLine(-50,25)(-25,25)
\ArrowLine(-25,25)(-5,25)
\ArrowLine(-25,-25)(-50,-25)
\ArrowLine(-5,-25)(-25,-25)
\Photon(-25,25)(-25,10){2}{3}
\ArrowArc(-25,0)(10,-90,90)
\ArrowArc(-25,0)(10,90,270)
\Photon(-25,-10)(-25,-25){2}{3} 
\Text(-22,-2)[cb]{{\small $2 J_{14} $}}
\end{picture}}
}
\hspace{-0.1cm}
\times
\hspace{1.4cm}
  \vcenter{
\hbox{\begin{picture}(0,0)(0,0)
\SetScale{.8}
  \SetWidth{.5}
\ArrowLine(-50,25)(-25,25)
\ArrowLine(-25,25)(-5,25)
\ArrowLine(-25,-25)(-50,-25)
\ArrowLine(-5,-25)(-25,-25)
\Photon(-25,25)(-25,-25){2}{10}
\end{picture}}
}
=
\mbox{IR fin.}
\]
\vspace*{.6cm}
\[\vcenter{
\hbox{
  \begin{picture}(0,0)(0,0)
\SetScale{.8}
  \SetWidth{.5}
\ArrowLine(-30,25)(-15,0) 
\ArrowLine(-15,0)(-30,-25) 
\ArrowLine(15,0)(30,25) 
\ArrowLine(30,-25)(15,0) 
\Photon(-15,0)(-10,0){2}{1.5}
\Photon(15,0)(10,0){2}{1.5}
\ArrowArc(0,0)(10,0,180)
\ArrowArc(0,0)(10,180,360)   
\end{picture}}  
}
\hspace{1cm}
\times
\hspace{1.4cm}
\vcenter{\hbox{\begin{picture}(0,0)(0,0)
\SetScale{.8}
  \SetWidth{.5}
\ArrowLine(-50,25)(-25,25)
\ArrowLine(-25,25)(25,25)
\ArrowLine(25,25)(50,25)
\Photon(-25,25)(25,-25){2}{10}
\Photon(25,25)(-25,-25){2}{10}
\ArrowLine(-25,-25)(-50,-25)
\ArrowLine(25,-25)(-25,-25)
\ArrowLine(50,-25)(25,-25)
\end{picture}
  }}
\hspace{1.5cm}
+
\hspace{3cm}
  \vcenter{
\hbox{\begin{picture}(0,0)(0,0)
\SetScale{.8}
  \SetWidth{.5}
\ArrowLine(-30,25)(-15,0) 
\ArrowLine(-15,0)(-30,-25) 
\ArrowLine(15,0)(30,25) 
\ArrowLine(30,-25)(15,0) 
\Photon(-15,0)(-10,0){2}{1.5}
\Photon(15,0)(10,0){2}{1.5}
\ArrowArc(0,0)(10,0,180)
\ArrowArc(0,0)(10,180,360)   
\Text(-18,-2)[cb]{{\small $2  J_{14} $}}
\end{picture}}
}
\hspace{1cm}
\times
\hspace{1.4cm}
  \vcenter{
\hbox{\begin{picture}(0,0)(0,0)
\SetScale{.8}
  \SetWidth{.5}
\ArrowLine(-50,25)(-25,25)
\ArrowLine(-25,25)(-5,25)
\ArrowLine(-25,-25)(-50,-25)
\ArrowLine(-5,-25)(-25,-25)
\Photon(-25,25)(-25,-25){2}{10}
\end{picture}}
}
=
\mbox{IR fin.}
\]
\vspace*{.6cm}
\[\vcenter{
\hbox{
  \begin{picture}(0,0)(0,0)
\SetScale{.8}
  \SetWidth{.5}
  \ArrowLine(-50,25)(-25,25)
\ArrowLine(-25,25)(-5,25)
\ArrowLine(-25,-25)(-50,-25)
\ArrowLine(-5,-25)(-25,-25)
\Photon(-25,25)(-25,10){2}{3}
\ArrowArc(-25,0)(10,-90,90)
\ArrowArc(-25,0)(10,90,270)
\Photon(-25,-10)(-25,-25){2}{3}
\end{picture}}  
}
\hspace{0.1cm}
\times
\hspace{1.4cm}
\vcenter{\hbox{
\begin{picture}(0,0)(0,0)
\SetScale{.8}
\SetWidth{.5}
\ArrowLine(-50,25)(-25,25)
\ArrowLine(25,25)(50,25)
\Photon(-25,-25)(25,-25){2}{10}
\Photon(-25,25)(25,25){2}{10}
\ArrowLine(-25,-25)(-50,-25)
\ArrowLine(50,-25)(25,-25)
\ArrowLine(-25,25)(-25,-25)
\ArrowLine(25,-25)(25,25)
\end{picture}
  }}
\hspace{1.5cm}
+
\hspace{3.1cm}
  \vcenter{
\hbox{\begin{picture}(0,0)(0,0)
\SetScale{.8}
  \SetWidth{.5}
\ArrowLine(-50,25)(-25,25)
\ArrowLine(-25,25)(-5,25)
\ArrowLine(-25,-25)(-50,-25)
\ArrowLine(-5,-25)(-25,-25)
\Photon(-25,25)(-25,10){2}{3}
\ArrowArc(-25,0)(10,-90,90)
\ArrowArc(-25,0)(10,90,270)
\Photon(-25,-10)(-25,-25){2}{3}
\Text(-20,-2)[cb]{{\small $2 J_{13} $}}
\end{picture}}
}
\hspace{-0.1cm}
\times
\hspace{1.1cm}
  \vcenter{
\hbox{\begin{picture}(0,0)(0,0)
\SetScale{.8}
  \SetWidth{.5}
\ArrowLine(-30,25)(-15,0) 
\ArrowLine(-15,0)(-30,-25) 
\ArrowLine(15,0)(30,25) 
\ArrowLine(30,-25)(15,0) 
\Photon(-15,0)(15,0){2}{6}
\end{picture}}
}
\hspace*{1cm}
=
\mbox{IR fin.}
\]
\vspace*{.6cm}
\[\vcenter{
\hbox{\begin{picture}(0,0)(0,0)
\SetScale{.8}
  \SetWidth{.5}
\ArrowLine(-30,25)(-15,0) 
\ArrowLine(-15,0)(-30,-25) 
\ArrowLine(15,0)(30,25) 
\ArrowLine(30,-25)(15,0) 
\Photon(-15,0)(-10,0){2}{1.5}
\Photon(15,0)(10,0){2}{1.5}
\ArrowArc(0,0)(10,0,180)
\ArrowArc(0,0)(10,180,360)
\end{picture}}
}
\hspace*{.8cm}
\times
\hspace*{1.2cm}
\vcenter{\hbox{
\begin{picture}(0,0)(0,0)
\SetScale{.8}
\SetWidth{.5}
\ArrowLine(-50,25)(-25,25)
\ArrowLine(25,25)(50,25)
\Photon(-25,-25)(25,-25){2}{10}
\Photon(-25,25)(25,25){2}{10}
\ArrowLine(-25,-25)(-50,-25)
\ArrowLine(50,-25)(25,-25)
\ArrowLine(-25,25)(-25,-25)
\ArrowLine(25,-25)(25,25)
\end{picture}
  }}
\hspace*{1.3cm}
+
\hspace*{3cm}
  \vcenter{
\hbox{\begin{picture}(0,0)(0,0)
\SetScale{.8}
  \SetWidth{.5}
\ArrowLine(-30,25)(-15,0) 
\ArrowLine(-15,0)(-30,-25) 
\ArrowLine(15,0)(30,25) 
\ArrowLine(30,-25)(15,0) 
\Photon(-15,0)(-10,0){2}{1.5}
\Photon(15,0)(10,0){2}{1.5}
\ArrowArc(0,0)(10,0,180)
\ArrowArc(0,0)(10,180,360)
\Text(-20,-2)[cb]{{\small $2 J_{13} $}}
\end{picture}}
}
\hspace{1cm}
\times
\hspace{1.1cm}
  \vcenter{
\hbox{\begin{picture}(0,0)(0,0)
\SetScale{.8}
  \SetWidth{.5}
\ArrowLine(-30,25)(-15,0) 
\ArrowLine(-15,0)(-30,-25) 
\ArrowLine(15,0)(30,25) 
\ArrowLine(30,-25)(15,0) 
\Photon(-15,0)(15,0){2}{6}
\end{picture}}
}
\hspace*{1cm}
=
\mbox{IR fin.}
\]
\vspace*{.6cm}
\[\vcenter{
\hbox{
  \begin{picture}(0,0)(0,0)
\SetScale{.8}
  \SetWidth{.5}
\ArrowLine(-50,25)(-25,25)
\ArrowLine(-25,25)(-5,25)
\ArrowLine(-25,-25)(-50,-25)
\ArrowLine(-5,-25)(-25,-25)
\Photon(-25,25)(-25,10){2}{3}
\ArrowArc(-25,0)(10,-90,90)
\ArrowArc(-25,0)(10,90,270)
\Photon(-25,-10)(-25,-25){2}{3}
\end{picture}}  
}
\hspace{0.1cm}
\times
\hspace{1.4cm}
\vcenter{\hbox{\begin{picture}(0,0)(0,0)
\SetScale{.8}
  \SetWidth{.5}
\ArrowLine(-50,25)(-25,25)
\ArrowLine(25,25)(50,25)
\Photon(-25,25)(25,-25){2}{10}
\Photon(-25,-25)(25,25){2}{10}
\ArrowLine(-25,-25)(-50,-25)
\ArrowLine(50,-25)(25,-25)
\ArrowLine(-25,25)(-25,-25)
\ArrowLine(25,-25)(25,25)
\end{picture}
  }}
\hspace{1.5cm}
+
\hspace{3.2cm}
  \vcenter{
\hbox{\begin{picture}(0,0)(0,0)
\SetScale{.8}
  \SetWidth{.5}
\ArrowLine(-50,25)(-25,25)
\ArrowLine(-25,25)(-5,25)
\ArrowLine(-25,-25)(-50,-25)
\ArrowLine(-5,-25)(-25,-25)
\Photon(-25,25)(-25,10){2}{3}
\ArrowArc(-25,0)(10,-90,90)
\ArrowArc(-25,0)(10,90,270)
\Photon(-25,-10)(-25,-25){2}{3}  
\Text(-23,-2)[cb]{{\small $2 J_{14}$}}
\end{picture}}
}
\hspace{-0.1cm}
\times
\hspace{1.1cm}
  \vcenter{
\hbox{\begin{picture}(0,0)(0,0)
\SetScale{.8}
  \SetWidth{.5}
\ArrowLine(-30,25)(-15,0) 
\ArrowLine(-15,0)(-30,-25) 
\ArrowLine(15,0)(30,25) 
\ArrowLine(30,-25)(15,0) 
\Photon(-15,0)(15,0){2}{6}
\end{picture}}
}
\hspace*{1cm}
=
\mbox{IR fin.}
\]
\vspace*{.6cm}
\[\vcenter{
\hbox{\begin{picture}(0,0)(0,0)
\SetScale{.8}
  \SetWidth{.5}
 \ArrowLine(-30,25)(-15,0) 
\ArrowLine(-15,0)(-30,-25) 
\ArrowLine(15,0)(30,25) 
\ArrowLine(30,-25)(15,0) 
\Photon(-15,0)(-10,0){2}{1.5}
\Photon(15,0)(10,0){2}{1.5}
\ArrowArc(0,0)(10,0,180)
\ArrowArc(0,0)(10,180,360) 
\end{picture}}
}
\hspace*{.8cm}
\times
\hspace*{1.2cm}
\vcenter{\hbox{\begin{picture}(0,0)(0,0)
\SetScale{.8}
  \SetWidth{.5}
\ArrowLine(-50,25)(-25,25)
\ArrowLine(25,25)(50,25)
\Photon(-25,25)(25,-25){2}{10}
\Photon(-25,-25)(25,25){2}{10}
\ArrowLine(-25,-25)(-50,-25)
\ArrowLine(50,-25)(25,-25)
\ArrowLine(-25,25)(-25,-25)
\ArrowLine(25,-25)(25,25)
\end{picture}
  }}
\hspace*{1.3cm}
+ 
\hspace*{3cm}
  \vcenter{
\hbox{\begin{picture}(0,0)(0,0)
\SetScale{.8}
  \SetWidth{.5}\ArrowLine(-30,25)(-15,0) 
\ArrowLine(-15,0)(-30,-25) 
\ArrowLine(15,0)(30,25) 
\ArrowLine(30,-25)(15,0) 
\Photon(-15,0)(-10,0){2}{1.5}
\Photon(15,0)(10,0){2}{1.5}
\ArrowArc(0,0)(10,0,180)
\ArrowArc(0,0)(10,180,360)
\Text(-20,-2)[cb]{{\small $2 J_{14} $}}
\end{picture}}
}
\hspace{1cm}
\times
\hspace{1.1cm}
  \vcenter{
\hbox{\begin{picture}(0,0)(0,0)
\SetScale{.8}
  \SetWidth{.5}
\ArrowLine(-30,25)(-15,0) 
\ArrowLine(-15,0)(-30,-25) 
\ArrowLine(15,0)(30,25) 
\ArrowLine(30,-25)(15,0) 
\Photon(-15,0)(15,0){2}{6}
\end{picture}}
}
\hspace*{1cm}
=
\mbox{IR fin.}
\]
\vspace*{.8cm}
\caption[]{\it Cancellation of the IR divergencies in the product of one-loop
box and self-energy diagrams.}
\label{1lboxtimes1lself}
\end{figure}
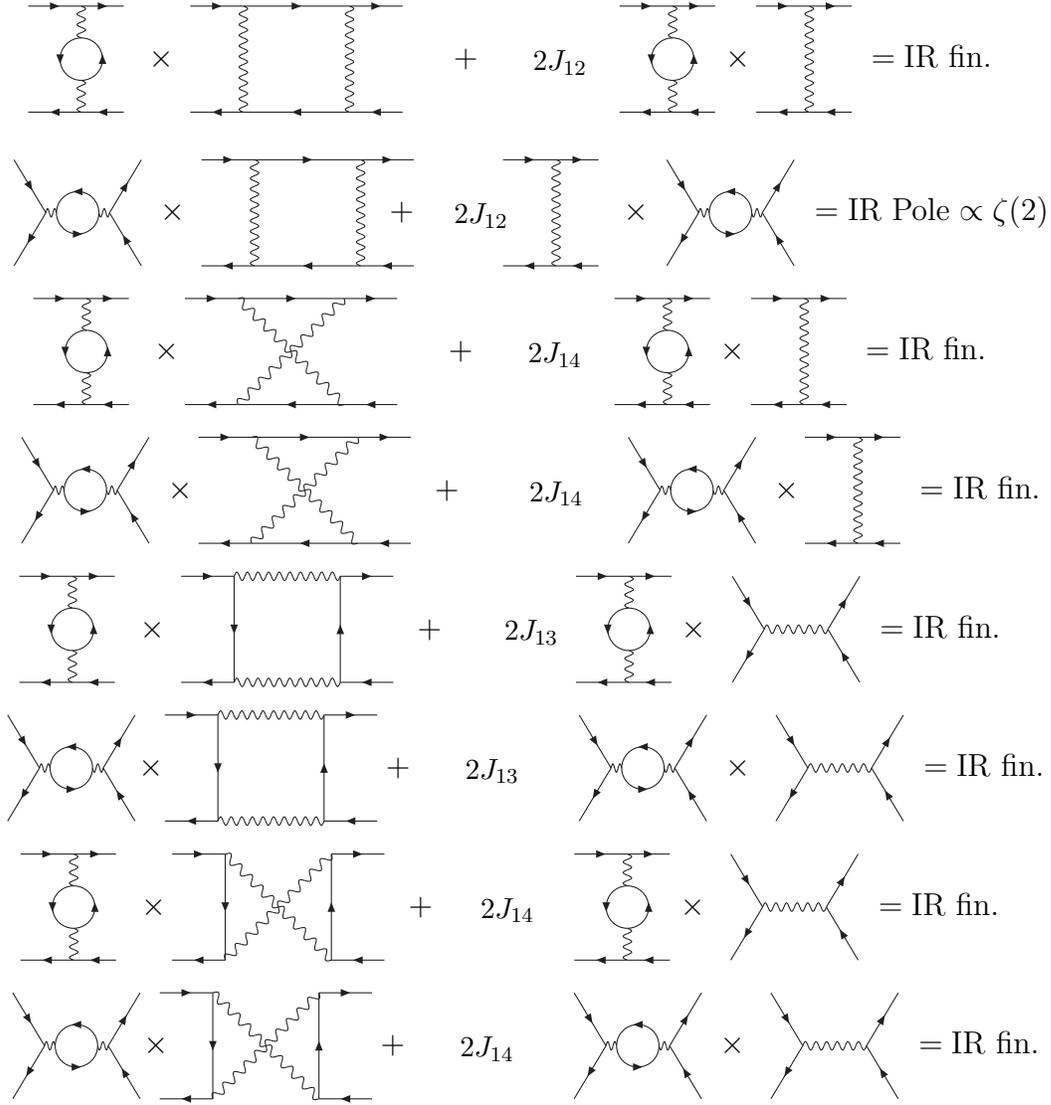
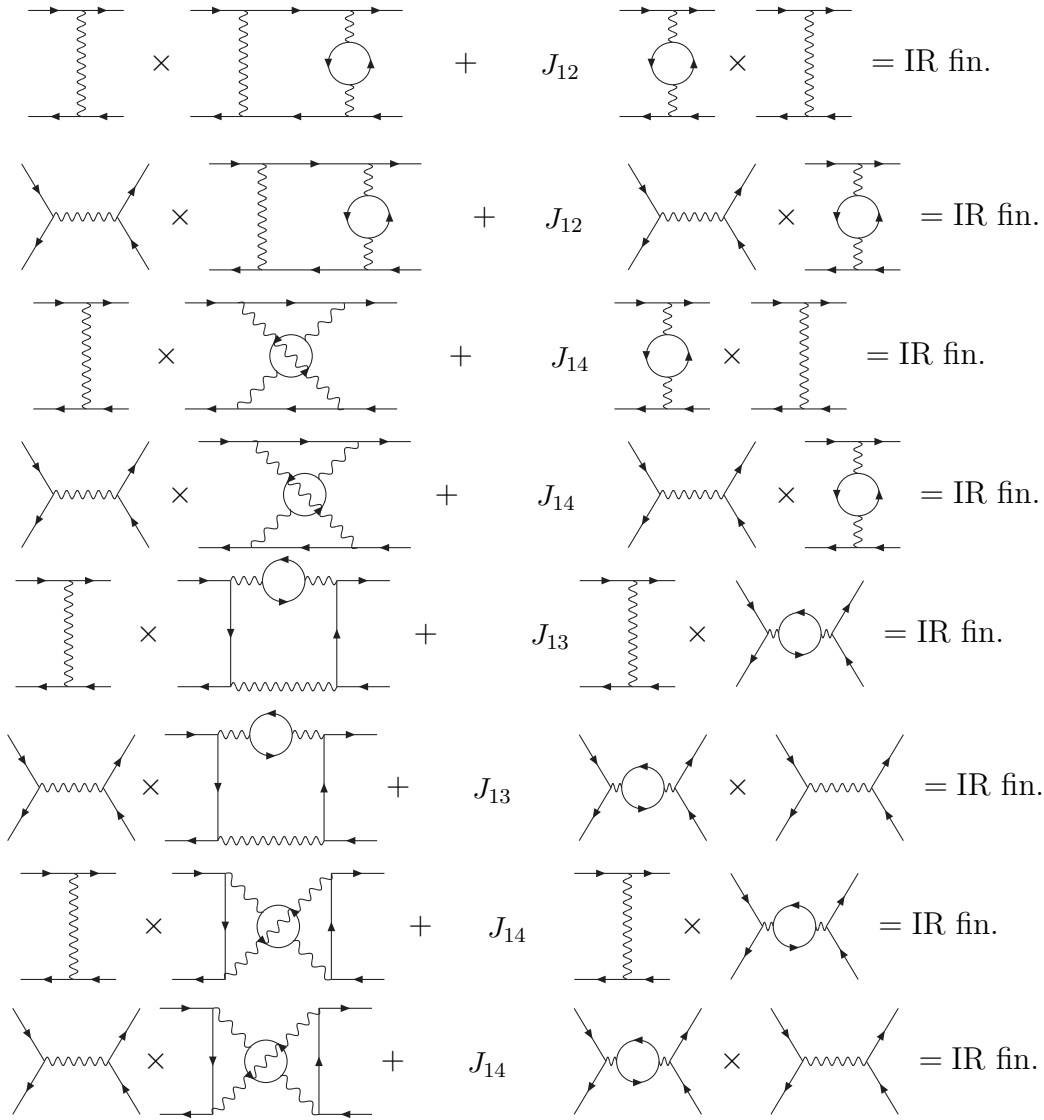
\begin{figure}
\vspace*{.6cm}
\[\vcenter{
\hbox{
  \begin{picture}(0,0)(0,0)
\SetScale{.8}
  \SetWidth{.5}
\ArrowLine(-50,25)(-25,25)
\ArrowLine(-25,25)(-5,25)
\ArrowLine(-25,-25)(-50,-25)
\ArrowLine(-5,-25)(-25,-25)
\Photon(-25,25)(-25,-25){2}{10}
\end{picture}}  
}
\hspace{0.1cm}
\times
\hspace{1.4cm}
\vcenter{\hbox{
  \begin{picture}(0,0)(0,0)
\SetScale{.8}
  \SetWidth{.5}
\ArrowLine(-50,25)(-25,25)
\ArrowLine(-25,25)(25,25)
\ArrowLine(25,25)(50,25)
\Photon(-25,25)(-25,-25){2}{10}
\Photon(25,25)(25,10){2}{3}
\Photon(25,-10)(25,-25){2}{3}
\ArrowArc(25,0)(10,-90,90)
\ArrowArc(25,0)(10,90,270)
\ArrowLine(-25,-25)(-50,-25)
\ArrowLine(25,-25)(-25,-25)
\ArrowLine(50,-25)(25,-25)
\end{picture}}}
\hspace{2cm}
+
\hspace{3.2cm}
  \vcenter{
\hbox{\begin{picture}(0,0)(0,0)
\SetScale{.8}
  \SetWidth{.5}
  \ArrowLine(-50,25)(-25,25)
\ArrowLine(-25,25)(-5,25)
\ArrowLine(-25,-25)(-50,-25)
\ArrowLine(-5,-25)(-25,-25)
\Photon(-25,25)(-25,10){2}{3}
\ArrowArc(-25,0)(10,-90,90)
\ArrowArc(-25,0)(10,90,270)
\Photon(-25,-10)(-25,-25){2}{3}
\Text(-22,-2)[cb]{{\small $ J_{12}  $}}
\end{picture}}
}
\hspace{-0.1cm}
\times
\hspace{1.4cm}
  \vcenter{
\hbox{\begin{picture}(0,0)(0,0)
\SetScale{.8}
  \SetWidth{.5}
\ArrowLine(-50,25)(-25,25)
\ArrowLine(-25,25)(-5,25)
\ArrowLine(-25,-25)(-50,-25)
\ArrowLine(-5,-25)(-25,-25)
\Photon(-25,25)(-25,-25){2}{10}
\end{picture}}
}
=
\mbox{IR fin.}
\]
\vspace*{.6cm}
\[\vcenter{
\hbox{
  \begin{picture}(0,0)(0,0)
\SetScale{.8}
  \SetWidth{.5}
\ArrowLine(-30,25)(-15,0) 
\ArrowLine(-15,0)(-30,-25) 
\ArrowLine(15,0)(30,25) 
\ArrowLine(30,-25)(15,0) 
\Photon(-15,0)(15,0){2}{6}
\end{picture}}  
}
\hspace{1cm}
\times
\hspace{1.4cm}
\vcenter{\hbox{
  \begin{picture}(0,0)(0,0)
\SetScale{.8}
  \SetWidth{.5}
\ArrowLine(-50,25)(-25,25)
\ArrowLine(-25,25)(25,25)
\ArrowLine(25,25)(50,25)
\Photon(-25,25)(-25,-25){2}{10}
\Photon(25,25)(25,10){2}{3}
\Photon(25,-10)(25,-25){2}{3}
\ArrowArc(25,0)(10,-90,90)
\ArrowArc(25,0)(10,90,270)
\ArrowLine(-25,-25)(-50,-25)
\ArrowLine(25,-25)(-25,-25)
\ArrowLine(50,-25)(25,-25)  
\end{picture}}}
\hspace{2cm}
+
\hspace{2.5cm}
  \vcenter{
\hbox{\begin{picture}(0,0)(0,0)
\SetScale{.8}
  \SetWidth{.5}
\ArrowLine(-30,25)(-15,0) 
\ArrowLine(-15,0)(-30,-25) 
\ArrowLine(15,0)(30,25) 
\ArrowLine(30,-25)(15,0) 
\Photon(-15,0)(15,0){2}{6}
\Text(-17,-2)[cb]{{\small $ J_{12} $}}
\end{picture}}
}
\hspace{1cm}
\times
\hspace{1.4cm}
  \vcenter{
\hbox{\begin{picture}(0,0)(0,0)
\SetScale{.8}
  \SetWidth{.5}
\ArrowLine(-50,25)(-25,25)
\ArrowLine(-25,25)(-5,25)
\ArrowLine(-25,-25)(-50,-25)
\ArrowLine(-5,-25)(-25,-25)
\Photon(-25,25)(-25,10){2}{3}
\ArrowArc(-25,0)(10,-90,90)
\ArrowArc(-25,0)(10,90,270)
\Photon(-25,-10)(-25,-25){2}{3}  
\end{picture}}
}
=
\mbox{IR fin.}
\]
\vspace*{.6cm}
\[\vcenter{
\hbox{
  \begin{picture}(0,0)(0,0)
\SetScale{.8}
  \SetWidth{.5}
\ArrowLine(-50,25)(-25,25)
\ArrowLine(-25,25)(-5,25)
\ArrowLine(-25,-25)(-50,-25)
\ArrowLine(-5,-25)(-25,-25)
\Photon(-25,25)(-25,-25){2}{10}
\end{picture}}  
}
\hspace{0.1cm}
\times
\hspace{1.4cm}
\vcenter{\hbox{\begin{picture}(0,0)(0,0)
\SetScale{.8}
  \SetWidth{.5}
\ArrowLine(-50,25)(-25,25)
\ArrowLine(-25,25)(25,25)
\ArrowLine(25,25)(50,25)
\Photon(-25,25)(25,-25){2}{10}
\Photon(25,25)(6.85,6.85){2}{3}
\Photon(-6.85,-6.85)(-25,-25){2}{3}
\ArrowArc(0,0)(10,-135,45)
\ArrowArc(0,0)(10,45,225)
\ArrowLine(-25,-25)(-50,-25)
\ArrowLine(25,-25)(-25,-25)
\ArrowLine(50,-25)(25,-25)
\end{picture}}}
\hspace{2cm}
+
\hspace{3.2cm}
  \vcenter{
\hbox{\begin{picture}(0,0)(0,0)
\SetScale{.8}
  \SetWidth{.5}
  \ArrowLine(-50,25)(-25,25)
\ArrowLine(-25,25)(-5,25)
\ArrowLine(-25,-25)(-50,-25)
\ArrowLine(-5,-25)(-25,-25)
\Photon(-25,25)(-25,10){2}{3}
\ArrowArc(-25,0)(10,-90,90)
\ArrowArc(-25,0)(10,90,270)
\Photon(-25,-10)(-25,-25){2}{3}
\Text(-20,-2)[cb]{{\small $ J_{14} $}}
\end{picture}}
}
\hspace{-0.1cm}
\times
\hspace{1.4cm}
  \vcenter{
\hbox{\begin{picture}(0,0)(0,0)
\SetScale{.8}
  \SetWidth{.5}
\ArrowLine(-50,25)(-25,25)
\ArrowLine(-25,25)(-5,25)
\ArrowLine(-25,-25)(-50,-25)
\ArrowLine(-5,-25)(-25,-25)
\Photon(-25,25)(-25,-25){2}{10}
\end{picture}}
}
=
\mbox{IR fin.}
\]
\vspace*{.6cm}
\[\vcenter{
\hbox{
  \begin{picture}(0,0)(0,0)
\SetScale{.8}
  \SetWidth{.5}
\ArrowLine(-30,25)(-15,0) 
\ArrowLine(-15,0)(-30,-25) 
\ArrowLine(15,0)(30,25) 
\ArrowLine(30,-25)(15,0) 
\Photon(-15,0)(15,0){2}{6}
\end{picture}}  
}
\hspace{1cm}
\times
\hspace{1.4cm}
\vcenter{\hbox{\begin{picture}(0,0)(0,0)
\SetScale{.8}
  \SetWidth{.5}
\ArrowLine(-50,25)(-25,25)
\ArrowLine(-25,25)(25,25)
\ArrowLine(25,25)(50,25)
\Photon(-25,25)(25,-25){2}{10}
\Photon(25,25)(6.85,6.85){2}{3}
\Photon(-6.85,-6.85)(-25,-25){2}{3}
\ArrowArc(0,0)(10,-135,45)
\ArrowArc(0,0)(10,45,225)
\ArrowLine(-25,-25)(-50,-25)
\ArrowLine(25,-25)(-25,-25)
\ArrowLine(50,-25)(25,-25)
\end{picture}
  }}
\hspace{1.5cm}
+
\hspace{3cm}
  \vcenter{
\hbox{\begin{picture}(0,0)(0,0)
\SetScale{.8}
  \SetWidth{.5}
\ArrowLine(-30,25)(-15,0) 
\ArrowLine(-15,0)(-30,-25) 
\ArrowLine(15,0)(30,25) 
\ArrowLine(30,-25)(15,0) 
\Photon(-15,0)(15,0){2}{6}
\Text(-18,-2)[cb]{{\small $ J_{14} $}}
\end{picture}}
}
\hspace{1cm}
\times
\hspace{1.4cm}
  \vcenter{
\hbox{\begin{picture}(0,0)(0,0)
\SetScale{.8}
  \SetWidth{.5}
\ArrowLine(-50,25)(-25,25)
\ArrowLine(-25,25)(-5,25)
\ArrowLine(-25,-25)(-50,-25)
\ArrowLine(-5,-25)(-25,-25)
\Photon(-25,25)(-25,10){2}{3}
\ArrowArc(-25,0)(10,-90,90)
\ArrowArc(-25,0)(10,90,270)
\Photon(-25,-10)(-25,-25){2}{3}   
\end{picture}}
}
=
\mbox{IR fin.}
\]
\vspace*{.6cm}
\[\vcenter{
\hbox{
  \begin{picture}(0,0)(0,0)
\SetScale{.8}
  \SetWidth{.5}
\ArrowLine(-50,25)(-25,25)
\ArrowLine(-25,25)(-5,25)
\ArrowLine(-25,-25)(-50,-25)
\ArrowLine(-5,-25)(-25,-25)
\Photon(-25,25)(-25,-25){2}{10}
\end{picture}}  
}
\hspace{0.1cm}
\times
\hspace{1.4cm}
\vcenter{\hbox{
\begin{picture}(0,0)(0,0)
\SetScale{.8}
\SetWidth{.5}
\ArrowLine(-50,25)(-25,25)
\ArrowLine(25,25)(50,25)
\Photon(-25,-25)(25,-25){2}{10}
\Photon(-25,25)(-10,25){2}{3}
\Photon(10,25)(25,25){2}{3}
\ArrowArc(0,25)(10,0,180)
\ArrowArc(0,25)(10,180,360)
\ArrowLine(-25,-25)(-50,-25)
\ArrowLine(50,-25)(25,-25)
\ArrowLine(-25,25)(-25,-25)
\ArrowLine(25,-25)(25,25)
\end{picture}
  }}
\hspace{1.5cm}
+
\hspace{3.2cm}
  \vcenter{
\hbox{\begin{picture}(0,0)(0,0)
\SetScale{.8}
  \SetWidth{.5}
\ArrowLine(-50,25)(-25,25)
\ArrowLine(-25,25)(-5,25)
\ArrowLine(-25,-25)(-50,-25)
\ArrowLine(-5,-25)(-25,-25)
\Photon(-25,25)(-25,-25){2}{10}
\Text(-18,-2)[cb]{{\small $ J_{13} $}}
\end{picture}}
}
\hspace{-0.1cm}
\times
\hspace{1.1cm}
  \vcenter{
\hbox{\begin{picture}(0,0)(0,0)
\SetScale{.8}
  \SetWidth{.5}
\ArrowLine(-30,25)(-15,0) 
\ArrowLine(-15,0)(-30,-25) 
\ArrowLine(15,0)(30,25) 
\ArrowLine(30,-25)(15,0) 
\Photon(-15,0)(-10,0){2}{1.5}
\Photon(15,0)(10,0){2}{1.5}
\ArrowArc(0,0)(10,0,180)
\ArrowArc(0,0)(10,180,360)
\end{picture}}
}
\hspace*{1cm}
=
\mbox{IR fin.}
\]
\vspace*{.8cm}
\[\vcenter{
\hbox{\begin{picture}(0,0)(0,0)
\SetScale{.8}
  \SetWidth{.5}
\ArrowLine(-30,25)(-15,0) 
\ArrowLine(-15,0)(-30,-25) 
\ArrowLine(15,0)(30,25) 
\ArrowLine(30,-25)(15,0) 
\Photon(-15,0)(15,0){2}{6}
\end{picture}}
}
\hspace*{.8cm}
\times
\hspace*{1.2cm}
\vcenter{\hbox{
\begin{picture}(0,0)(0,0)
\SetScale{.8}
\SetWidth{.5}
\ArrowLine(-50,25)(-25,25)
\ArrowLine(25,25)(50,25)
\Photon(-25,-25)(25,-25){2}{10}
\Photon(-25,25)(-10,25){2}{3}
\Photon(10,25)(25,25){2}{3}
\ArrowArc(0,25)(10,0,180)
\ArrowArc(0,25)(10,180,360)
\ArrowLine(-25,-25)(-50,-25)
\ArrowLine(50,-25)(25,-25)
\ArrowLine(-25,25)(-25,-25)
\ArrowLine(25,-25)(25,25)
\end{picture}
  }}
\hspace*{1.3cm}
+
\hspace*{3cm}
  \vcenter{
\hbox{\begin{picture}(0,0)(0,0)
\SetScale{.8}
  \SetWidth{.5}
  \ArrowLine(-30,25)(-15,0) 
\ArrowLine(-15,0)(-30,-25) 
\ArrowLine(15,0)(30,25) 
\ArrowLine(30,-25)(15,0) 
\Photon(-15,0)(-10,0){2}{1.5}
\Photon(15,0)(10,0){2}{1.5}
\ArrowArc(0,0)(10,0,180)
\ArrowArc(0,0)(10,180,360)
\Text(-20,-2)[cb]{{\small $ J_{13} $}}
\end{picture}}
}
\hspace{1cm}
\times
\hspace{1.1cm}
  \vcenter{
\hbox{\begin{picture}(0,0)(0,0)
\SetScale{.8}
  \SetWidth{.5}
\ArrowLine(-30,25)(-15,0) 
\ArrowLine(-15,0)(-30,-25) 
\ArrowLine(15,0)(30,25) 
\ArrowLine(30,-25)(15,0) 
\Photon(-15,0)(15,0){2}{6}
\end{picture}}
}
\hspace*{1cm}
=
\mbox{IR fin.}
\]
\vspace*{.6cm}
\[\vcenter{
\hbox{
  \begin{picture}(0,0)(0,0)
\SetScale{.8}
  \SetWidth{.5}
\ArrowLine(-50,25)(-25,25)
\ArrowLine(-25,25)(-5,25)
\ArrowLine(-25,-25)(-50,-25)
\ArrowLine(-5,-25)(-25,-25)
\Photon(-25,25)(-25,-25){2}{10}
\end{picture}}  
}
\hspace{0.1cm}
\times
\hspace{1.4cm}
\vcenter{\hbox{\begin{picture}(0,0)(0,0)
\SetScale{.8}
  \SetWidth{.5}
\ArrowLine(-50,25)(-25,25)
\ArrowLine(25,25)(50,25)
\Photon(-25,25)(-6.85,6.85){2}{3}
\Photon(6.85,-6.85)(25,-25){2}{3}
\ArrowArc(0,0)(10,-45,135)
\ArrowArc(0,0)(10,135,315)
\Photon(-25,-25)(25,25){2}{10}
\ArrowLine(-25,-25)(-50,-25)
\ArrowLine(50,-25)(25,-25)
\ArrowLine(-25,25)(-25,-25)
\ArrowLine(25,-25)(25,25)
\end{picture}
  }}
\hspace{1.5cm}
+
\hspace{3.2cm}
  \vcenter{
\hbox{\begin{picture}(0,0)(0,0)
\SetScale{.8}
  \SetWidth{.5}
\ArrowLine(-50,25)(-25,25)
\ArrowLine(-25,25)(-5,25)
\ArrowLine(-25,-25)(-50,-25)
\ArrowLine(-5,-25)(-25,-25)
\Photon(-25,25)(-25,-25){2}{10}
\Text(-23,-2)[cb]{{\small $ J_{14}$}}
\end{picture}}
}
\hspace{-0.1cm}
\times
\hspace{1.1cm}
  \vcenter{
\hbox{\begin{picture}(0,0)(0,0)
\SetScale{.8}
  \SetWidth{.5}
  \ArrowLine(-30,25)(-15,0) 
\ArrowLine(-15,0)(-30,-25) 
\ArrowLine(15,0)(30,25) 
\ArrowLine(30,-25)(15,0) 
\Photon(-15,0)(-10,0){2}{1.5}
\Photon(15,0)(10,0){2}{1.5}
\ArrowArc(0,0)(10,0,180)
\ArrowArc(0,0)(10,180,360)
\end{picture}}
}
\hspace*{1cm}
=
\mbox{IR fin.}
\]

\vspace*{.6cm}
\[\vcenter{
\hbox{\begin{picture}(0,0)(0,0)
\SetScale{.8}
  \SetWidth{.5}
\ArrowLine(-30,25)(-15,0) 
\ArrowLine(-15,0)(-30,-25) 
\ArrowLine(15,0)(30,25) 
\ArrowLine(30,-25)(15,0) 
\Photon(-15,0)(15,0){2}{6}
\end{picture}}
}
\hspace*{.8cm}
\times
\hspace*{1.2cm}
\vcenter{\hbox{\begin{picture}(0,0)(0,0)
\SetScale{.8}
  \SetWidth{.5}
\ArrowLine(-50,25)(-25,25)
\ArrowLine(25,25)(50,25)
\Photon(-25,25)(-6.85,6.85){2}{3}
\Photon(6.85,-6.85)(25,-25){2}{3}
\ArrowArc(0,0)(10,-45,135)
\ArrowArc(0,0)(10,135,315)
\Photon(-25,-25)(25,25){2}{10}
\ArrowLine(-25,-25)(-50,-25)
\ArrowLine(50,-25)(25,-25)
\ArrowLine(-25,25)(-25,-25)
\ArrowLine(25,-25)(25,25)  
\end{picture}
  }}
\hspace*{1.3cm}
+ 
\hspace*{3cm}
  \vcenter{
\hbox{\begin{picture}(0,0)(0,0)
\SetScale{.8}
  \SetWidth{.5}
  \ArrowLine(-30,25)(-15,0) 
\ArrowLine(-15,0)(-30,-25) 
\ArrowLine(15,0)(30,25) 
\ArrowLine(30,-25)(15,0) 
\Photon(-15,0)(-10,0){2}{1.5}
\Photon(15,0)(10,0){2}{1.5}
\ArrowArc(0,0)(10,0,180)
\ArrowArc(0,0)(10,180,360)
\Text(-20,-2)[cb]{{\small $  J_{14} $}}
\end{picture}}
}
\hspace{1cm}
\times
\hspace{1.1cm}
  \vcenter{
\hbox{\begin{picture}(0,0)(0,0)
\SetScale{.8}
  \SetWidth{.5}
\ArrowLine(-30,25)(-15,0) 
\ArrowLine(-15,0)(-30,-25) 
\ArrowLine(15,0)(30,25) 
\ArrowLine(30,-25)(15,0) 
\Photon(-15,0)(15,0){2}{6}
\end{picture}}
}
\hspace*{1cm}
=
\mbox{IR fin.}
\]
\vspace*{.8cm}
\caption[]{\it Cancellation of the IR divergencies in the two-loop box diagrams.}
\label{2lboxes}
\end{figure}
where the integrals $J_{1j}$ ($j =1,\cdots,4$) have been introduced in the previous 
subsection.  The term proportional to $(D-4)$ in Eq.~(\ref{newV}) provides a
finite contribution to the real corrections in  Eq.~(\ref{above2}), since
$J_{1j}$ contains an IR pole. Terms proportional to $(D-4)$ in
$\sigma_2^S(s,t,m^2)$ are then neglected.

Figs.~\ref{2lreducible}-\ref{2lboxes} explicitly  show how the cancellation of
the IR divergences takes place from a diagrammatic point of view: the
contribution to the virtual cross-section of the interference of the diagrams
in the first term of each line is IR divergent; such divergence cancels against
the interference of the two diagrams in the second term multiplied by the
appropriate combination of $J_{1j}$ integrals.    We observe  that in the last
two lines of Fig.~\ref{2lreducible},  in the last line of
Fig.~\ref{onelVtimesonelS}, and in the second line of
Fig.~\ref{1lboxtimes1lself}, the subtraction of the real radiation in the
second term of the l.h.s. does not cancel completely the IR pole in the
corresponding virtual correction (first term in the l.h.s.). A residual IR
pole, proportional  to $\zeta(2)$, remains. As expected, the sum of the
residual poles vanishes and the cross-section is therefore IR finite.


\section{Numerical Results \label{Num}}

In order to  numerically evaluate the Bhabha scattering cross-section up to
corrections of order $\alpha^4 (N_F =1)$, we developed two computer 
codes. One of them was written in {\tt Mathematica} \cite{Mathematica}, while
the other was written in {\tt Fortran77}. Following~\cite{Bhabha1loopB}, in the
numerical calculation,  we fixed the energy cut-off on the undetected soft
photon to $\omega = 0.1 \, E$,  where $E$ is the beam energy. We compared the
results of the two codes  for a  beam energy ranging from  $E = 0.01 \,
\mbox{GeV}$ to $E = 500 \, \mbox{GeV}$, and    for an arbitrary choice of the
scattering angle $\theta$, finding complete  agreement.

\begin{figure} 
\bc
\begin{picture}(0,0)%
\includegraphics{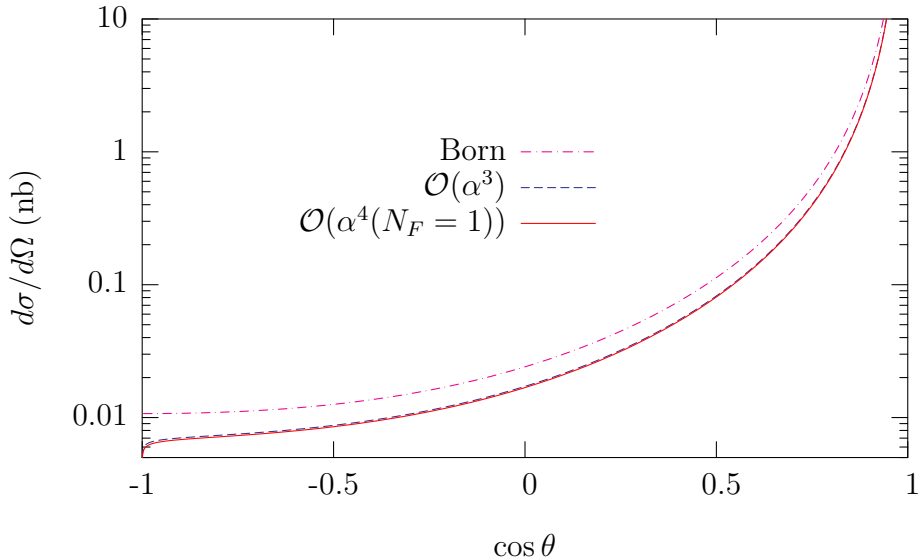}%
\end{picture}%
\setlength{\unitlength}{0.0200bp}%
\begin{picture}(18000,10800)(0,0)%
\put(2475,2734){\makebox(0,0)[r]{\strut{} 0.01}}%
\put(2475,5239){\makebox(0,0)[r]{\strut{} 0.1}}%
\put(2475,7745){\makebox(0,0)[r]{\strut{} 1}}%
\put(2475,10250){\makebox(0,0)[r]{\strut{} 10}}%
\put(2750,1430){\makebox(0,0){\strut{}-1}}%
\put(6356,1430){\makebox(0,0){\strut{}-0.5}}%
\put(9963,1430){\makebox(0,0){\strut{} 0}}%
\put(13569,1430){\makebox(0,0){\strut{} 0.5}}%
\put(17175,1430){\makebox(0,0){\strut{} 1}}%
\put(550,5950){\rotatebox{90}{\makebox(0,0){\strut{}$d\sigma/d\Omega$ (nb)}}}%
\put(9962,275){\makebox(0,0){\strut{}$\cos{\theta}$}}%
\put(9615,7745){\makebox(0,0)[r]{\strut{}Born}}%
\put(9615,7070){\makebox(0,0)[r]{\strut{}${\mathcal O}(\alpha^3)$}}%
\put(9615,6395){\makebox(0,0)[r]{\strut{}${\mathcal O}(\alpha^4(N_F=1))$}}%
\end{picture}
\caption{\it{Differential cross-section for $E = 22\, \mbox{GeV}$. The
dashed-dotted line corresponds to the Born cross-section, the dashed one to
the cross-section up to corrections of order $\alpha^3$, and the
continuous one to the  cross-section up to and including corrections of order
$\alpha^4(N_F=1)$.}} 
\label{CS22GEV} 
\ec
\end{figure}

In Fig.~\ref{CS22GEV}, we show the differential cross-section for $E= 22 \,
\mbox{GeV}$ and for $0 < \theta < \pi$.  The dashed-dotted line represents the 
cross-section in the Born approximation (Eq.~(\ref{BORN})), while the
continuous  (dashed) line represents the cross-section at order $\alpha^3$ 
($\alpha^4 (N_F =1)$), Eqs.~(\ref{1LOOP},\ref{2LOOP}). The radiative
corrections at order $\alpha^3$ are negative; therefore, they lower the 
cross-section. The corrections at order $\alpha^4 (N_F =1)$ are  negative and 
very small with respect to the corrections of order $\alpha^3$; they lower
the differential cross-section, although, as seen in  Fig.~\ref{CS22GEV}, the
effect is difficult to appreciate graphically.

\begin{figure}
\bc
\begin{picture}(0,0)%
\includegraphics{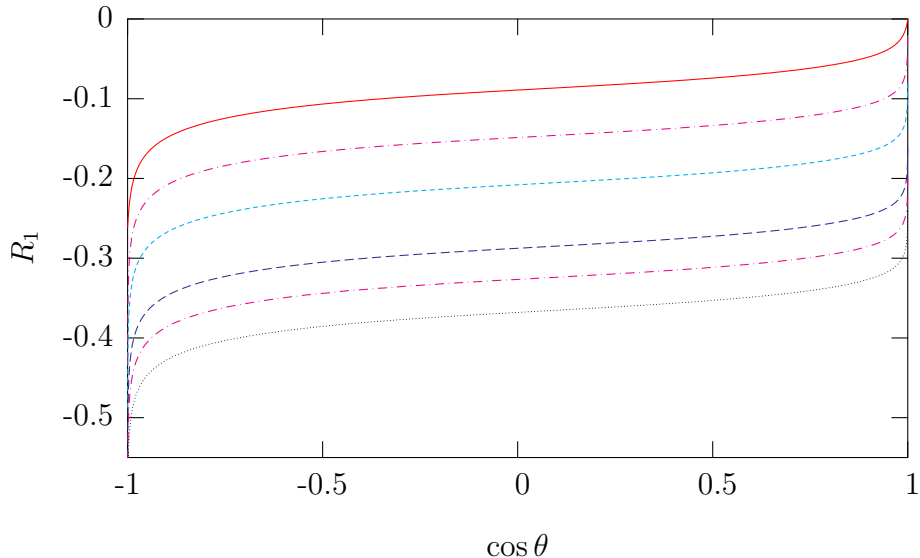}%
\end{picture}%
\setlength{\unitlength}{0.0200bp}%
\begin{picture}(18000,10800)(0,0)%
\put(2200,2732){\makebox(0,0)[r]{\strut{}-0.5}}%
\put(2200,4235){\makebox(0,0)[r]{\strut{}-0.4}}%
\put(2200,5739){\makebox(0,0)[r]{\strut{}-0.3}}%
\put(2200,7243){\makebox(0,0)[r]{\strut{}-0.2}}%
\put(2200,8746){\makebox(0,0)[r]{\strut{}-0.1}}%
\put(2200,10250){\makebox(0,0)[r]{\strut{} 0}}%
\put(2475,1430){\makebox(0,0){\strut{}-1}}%
\put(6150,1430){\makebox(0,0){\strut{}-0.5}}%
\put(9825,1430){\makebox(0,0){\strut{} 0}}%
\put(13500,1430){\makebox(0,0){\strut{} 0.5}}%
\put(17175,1430){\makebox(0,0){\strut{} 1}}%
\put(550,5950){\rotatebox{90}{\makebox(0,0){\strut{}$R_1$}}}%
\put(9825,275){\makebox(0,0){\strut{}$\cos{\theta}$}}%
\end{picture}
\caption{\it{Ratio $R_1$ of the one-loop corrections over the tree-level cross-section.
The different lines correspond to the following beam energies (from top to
bottom): $E = 10 \, {\rm MeV}$, $100$ {\rm MeV}, $1$ {\rm GeV}, $22$ {\rm GeV}, 
$100$ {\rm GeV}, and $500$ {\rm GeV}.}}
\label{RE1Ltot}
\ec
\end{figure}

\begin{figure}
\bc
\begin{picture}(0,0)%
\includegraphics{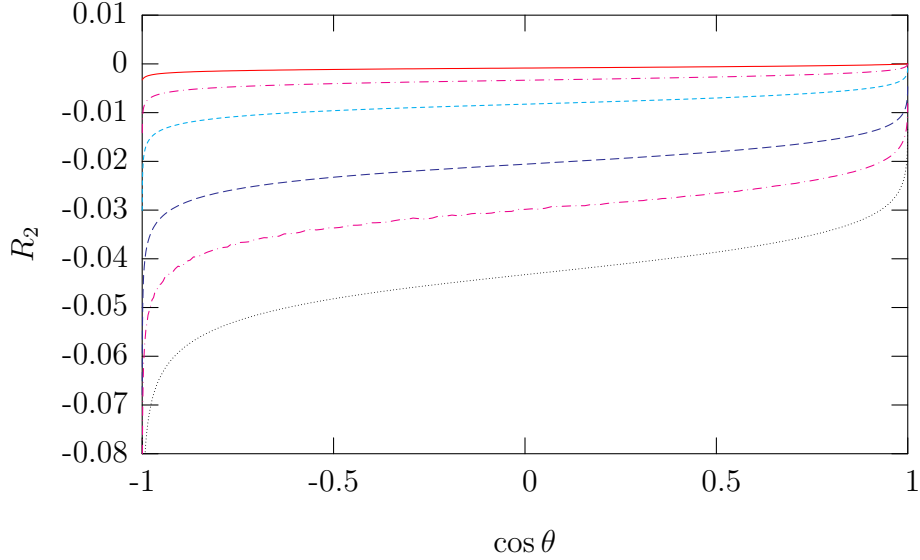}%
\end{picture}%
\setlength{\unitlength}{0.0200bp}%
\begin{picture}(18000,10800)(0,0)%
\put(2475,1980){\makebox(0,0)[r]{\strut{}-0.08}}%
\put(2475,2899){\makebox(0,0)[r]{\strut{}-0.07}}%
\put(2475,3818){\makebox(0,0)[r]{\strut{}-0.06}}%
\put(2475,4737){\makebox(0,0)[r]{\strut{}-0.05}}%
\put(2475,5656){\makebox(0,0)[r]{\strut{}-0.04}}%
\put(2475,6574){\makebox(0,0)[r]{\strut{}-0.03}}%
\put(2475,7493){\makebox(0,0)[r]{\strut{}-0.02}}%
\put(2475,8412){\makebox(0,0)[r]{\strut{}-0.01}}%
\put(2475,9331){\makebox(0,0)[r]{\strut{} 0}}%
\put(2475,10250){\makebox(0,0)[r]{\strut{} 0.01}}%
\put(2750,1430){\makebox(0,0){\strut{}-1}}%
\put(6356,1430){\makebox(0,0){\strut{}-0.5}}%
\put(9963,1430){\makebox(0,0){\strut{} 0}}%
\put(13569,1430){\makebox(0,0){\strut{} 0.5}}%
\put(17175,1430){\makebox(0,0){\strut{} 1}}%
\put(550,5950){\rotatebox{90}{\makebox(0,0){\strut{}$R_2$}}}%
\put(9962,275){\makebox(0,0){\strut{}$\cos{\theta}$}}%
\end{picture}
\caption{\it{Ratio $R_2$ of the two-loop corrections over the tree-level  
cross-section.The different lines correspond to the following beam energies 
(from top to bottom): $E = 10 \, {\rm MeV}$, $100$ {\rm MeV}, $1$ {\rm GeV}, 
$22$ {\rm GeV}, $100$ {\rm GeV}, and $500$ {\rm GeV}.}}
\label{RE2Ltot}
\ec
\end{figure}

The relative weight of the corrections of order $\alpha^3$ and  $\alpha^4 (N_F
=1)$ are shown in Fig.~\ref{RE1Ltot} and Fig.~\ref{RE2Ltot} for six different
choices of the beam energy. 
In Fig.~\ref{RE1Ltot}, we plotted the ratio of the
order $\alpha^3$ corrections to the tree-level cross-section:
\be
R_1 = \frac{\alpha}{\pi} \, \left( \frac{
d \sigma^{T}_1}{d \Omega} \right) \left(
\frac{d \sigma_0}{d \Omega} \right)^{-1} \, .
\ee
It is evident that the relative weight of the correction increases 
in magnitude with the beam energy and, at a given energy, with the scattering 
angle. A similar plot can be found in \cite{Bhabha1loopB}, for 
$E= 22 \, \mbox{GeV}$, where the full set of one-loop corrections in the Standard 
Model were considered. At $\theta = \pi/2$ the corrections range from $-9$\% of 
the Born cross-section at $E = 10 \, \mbox{MeV}$ to $-37$\% for 
$E = 500 \, \mbox{GeV}$.

In Fig.~\ref{RE2Ltot}, we plotted the ratio of the order $\alpha^4 (N_F =1)$
corrections to the complete order $\alpha^3$ cross-section:
\be
R_2 = \left( \frac{\alpha}{\pi} \right)^2  \left(
\frac{d \sigma^{T}_2}{d \Omega} \right) \left( \frac{d \sigma_0}{d \Omega} + 
\frac{\alpha}{\pi} \frac{d \sigma_1^{T}}{d \Omega} \right)^{-1} \, .
\ee
The corrections increase in magnitude with the energy and, for fixed energy,
with the scattering angle. The relative weight of these corrections 
at $\theta = \pi/2$ ranges from  $-0.08$\% of the complete cross-section at 
order $\alpha^3$ for $E = 10 \, \mbox{MeV}$ to $-4.3$\%  for 
$E = 500 \, \mbox{GeV}$.

\begin{figure}
\bc
\begin{picture}(0,0)%
\includegraphics{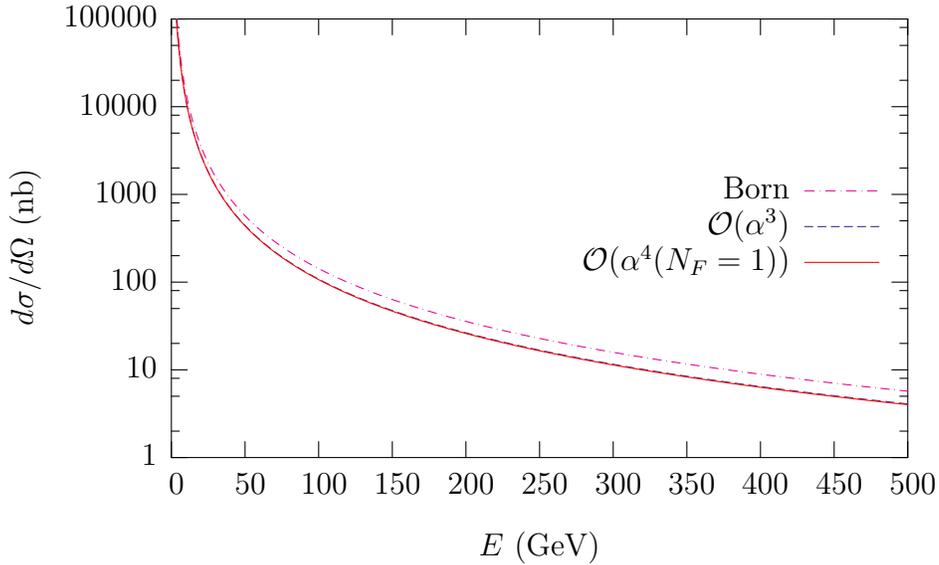}%
\end{picture}%
\setlength{\unitlength}{0.0200bp}%
\begin{picture}(18000,10800)(0,0)%
\put(3025,1980){\makebox(0,0)[r]{\strut{} 1}}%
\put(3025,3634){\makebox(0,0)[r]{\strut{} 10}}%
\put(3025,5288){\makebox(0,0)[r]{\strut{} 100}}%
\put(3025,6942){\makebox(0,0)[r]{\strut{} 1000}}%
\put(3025,8596){\makebox(0,0)[r]{\strut{} 10000}}%
\put(3025,10250){\makebox(0,0)[r]{\strut{} 100000}}%
\put(3300,1430){\makebox(0,0){\strut{} 0}}%
\put(4688,1430){\makebox(0,0){\strut{} 50}}%
\put(6075,1430){\makebox(0,0){\strut{} 100}}%
\put(7463,1430){\makebox(0,0){\strut{} 150}}%
\put(8850,1430){\makebox(0,0){\strut{} 200}}%
\put(10238,1430){\makebox(0,0){\strut{} 250}}%
\put(11625,1430){\makebox(0,0){\strut{} 300}}%
\put(13013,1430){\makebox(0,0){\strut{} 350}}%
\put(14400,1430){\makebox(0,0){\strut{} 400}}%
\put(15788,1430){\makebox(0,0){\strut{} 450}}%
\put(17175,1430){\makebox(0,0){\strut{} 500}}%
\put(550,5950){\rotatebox{90}{\makebox(0,0){\strut{}$d\sigma/d\Omega$ (nb)}}}%
\put(10237,275){\makebox(0,0){\strut{}$E$ (GeV)}}%
\put(14958,7010){\makebox(0,0)[r]{\strut{}Born}}%
\put(14958,6335){\makebox(0,0)[r]{\strut{}${\mathcal O}(\alpha^3)$}}%
\put(14958,5660){\makebox(0,0)[r]{\strut{}${\mathcal O}(\alpha^4(N_F=1))$}}%
\end{picture}
\caption{\it{Bhabha scattering cross-section as a function of the beam energy
for a scattering angle $\theta = 5^{\circ}$. The dashed-dotted line corresponds
to the Born cross-section, the dashed one to the cross-section up to
corrections of order $\alpha^3$, and the continuous one to the 
cross-section up to corrections of order $\alpha^4(N_F=1)$.}}
\label{CS5gr}
\ec
\end{figure}

\begin{figure}
\bc
\begin{picture}(0,0)%
\includegraphics{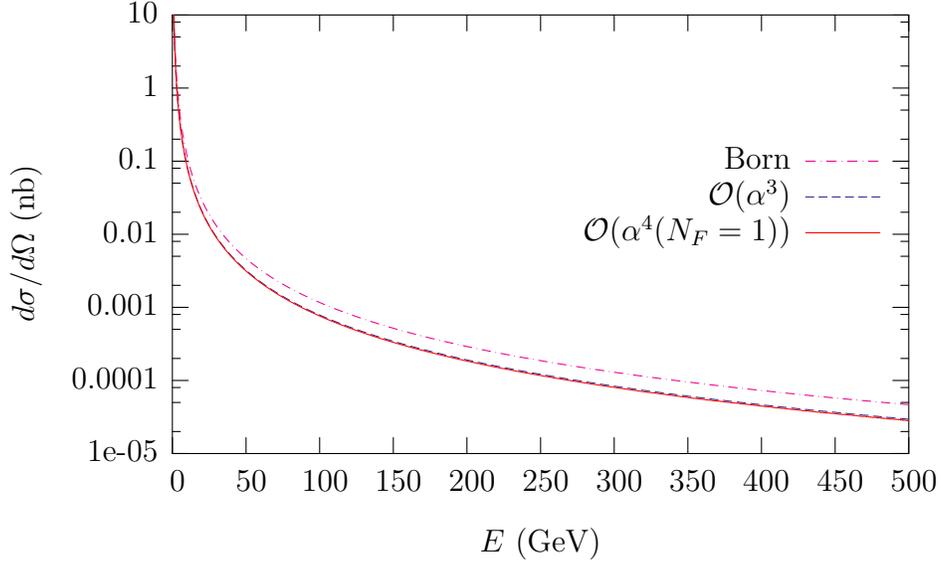}%
\end{picture}%
\setlength{\unitlength}{0.0200bp}%
\begin{picture}(18000,10800)(0,0)%
\put(3025,1980){\makebox(0,0)[r]{\strut{} 1e-05}}%
\put(3025,3358){\makebox(0,0)[r]{\strut{} 0.0001}}%
\put(3025,4737){\makebox(0,0)[r]{\strut{} 0.001}}%
\put(3025,6115){\makebox(0,0)[r]{\strut{} 0.01}}%
\put(3025,7493){\makebox(0,0)[r]{\strut{} 0.1}}%
\put(3025,8872){\makebox(0,0)[r]{\strut{} 1}}%
\put(3025,10250){\makebox(0,0)[r]{\strut{} 10}}%
\put(3300,1430){\makebox(0,0){\strut{} 0}}%
\put(4688,1430){\makebox(0,0){\strut{} 50}}%
\put(6075,1430){\makebox(0,0){\strut{} 100}}%
\put(7463,1430){\makebox(0,0){\strut{} 150}}%
\put(8850,1430){\makebox(0,0){\strut{} 200}}%
\put(10238,1430){\makebox(0,0){\strut{} 250}}%
\put(11625,1430){\makebox(0,0){\strut{} 300}}%
\put(13013,1430){\makebox(0,0){\strut{} 350}}%
\put(14400,1430){\makebox(0,0){\strut{} 400}}%
\put(15788,1430){\makebox(0,0){\strut{} 450}}%
\put(17175,1430){\makebox(0,0){\strut{} 500}}%
\put(550,5950){\rotatebox{90}{\makebox(0,0){\strut{}$d\sigma/d\Omega$ (nb)}}}%
\put(10237,275){\makebox(0,0){\strut{}$E$ (GeV)}}%
\put(14958,7493){\makebox(0,0)[r]{\strut{}Born}}%
\put(14958,6818){\makebox(0,0)[r]{\strut{}${\mathcal O}(\alpha^3)$}}%
\put(14958,6143){\makebox(0,0)[r]{\strut{}${\mathcal O}(\alpha^4(N_F=1))$}}%
\end{picture}
\caption{\it{Bhabha scattering cross-section as a function of the beam energy
for a scattering angle $\theta = 90^{\circ}$. The dashed-dotted line corresponds
to the Born cross-section, the dashed one to the cross-section up to
corrections of order $\alpha^3$, and the continuous one to the 
cross-section up to corrections of order $\alpha^4(N_F=1)$.}}
\label{CS90gr}
\ec
\end{figure}

Figs.~\ref{CS5gr} and~\ref{CS90gr} show the dependence of the cross-section
on the beam energy, for small and large scattering angles respectively.
Similar plots for a wider choices of angles can be found, limited to the
one-loop corrections, in \cite{Bhabha1loopC}. Using our codes, we reproduced
the plots shown in \cite{Bhabha1loopC} finding agreement.

To complete the analysis of the numerical results, we expanded the analytic
expression of the cross-section in the limit in which the squared electron mass
is negligible with respect to the kinematic invariants $s$, $t$, and $u$. 
We define the leading terms of the cross-section in the limit $m^2 \to 0$
through the relations:
\be
\frac{d \sigma^{T}_i}{d \Omega} = \left. \frac{d \sigma^{T}_i}{d \Omega}
\right|_L + {\mathcal O} \left( \frac{m^2}{s}, \frac{m^2}{t}, \frac{m^2}{u}
\right) \, , 
\label{LL}
\ee
where $i=1,2$ and where the subscript ``$L$'' stands for ``leading''.
The expressions of $(d \sigma^{T}_i/d \Omega)|_L$ can be found in 
Appendix~\ref{expandedres}.

\begin{figure}
\bc
\begin{picture}(0,0)%
\includegraphics{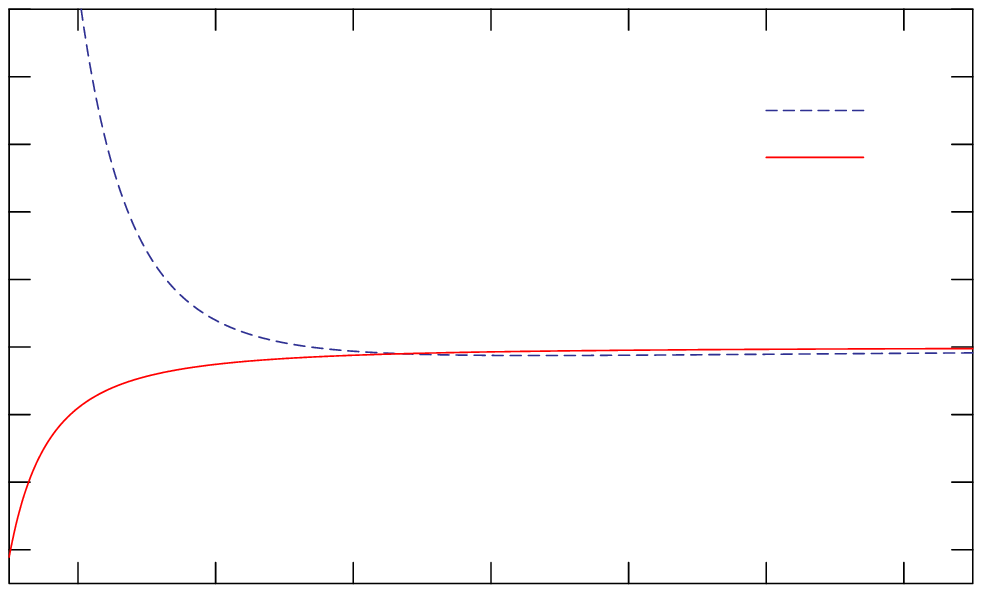}%
\end{picture}%
\setlength{\unitlength}{0.0200bp}%
\begin{picture}(18000,10800)(0,0)%
\put(3025,2466){\makebox(0,0)[r]{\strut{}-0.0006}}%
\put(3025,3439){\makebox(0,0)[r]{\strut{}-0.0004}}%
\put(3025,4412){\makebox(0,0)[r]{\strut{}-0.0002}}%
\put(3025,5385){\makebox(0,0)[r]{\strut{} 0}}%
\put(3025,6358){\makebox(0,0)[r]{\strut{} 0.0002}}%
\put(3025,7331){\makebox(0,0)[r]{\strut{} 0.0004}}%
\put(3025,8304){\makebox(0,0)[r]{\strut{} 0.0006}}%
\put(3025,9277){\makebox(0,0)[r]{\strut{} 0.0008}}%
\put(3025,10250){\makebox(0,0)[r]{\strut{} 0.001}}%
\put(4291,1430){\makebox(0,0){\strut{} 0.02}}%
\put(6273,1430){\makebox(0,0){\strut{} 0.04}}%
\put(8255,1430){\makebox(0,0){\strut{} 0.06}}%
\put(10238,1430){\makebox(0,0){\strut{} 0.08}}%
\put(12220,1430){\makebox(0,0){\strut{} 0.1}}%
\put(14202,1430){\makebox(0,0){\strut{} 0.12}}%
\put(16184,1430){\makebox(0,0){\strut{} 0.14}}%
\put(550,5950){\rotatebox{90}{\makebox(0,0){\strut{}$D_1$}}}%
\put(10237,275){\makebox(0,0){\strut{}$E$ (GeV)}}%
\put(13927,8791){\makebox(0,0)[r]{\strut{}$5^{\circ}$}}%
\put(13927,8116){\makebox(0,0)[r]{\strut{}$90^{\circ}$}}%
\end{picture}
\caption{\it{$D_1$ for $\theta = 5^{\circ}$ (dashed line) and $90^{\circ}$
(continuous line).}}
\label{DIFF1l}
\ec
\end{figure}

\begin{figure}
\bc
\begin{picture}(0,0)%
\includegraphics{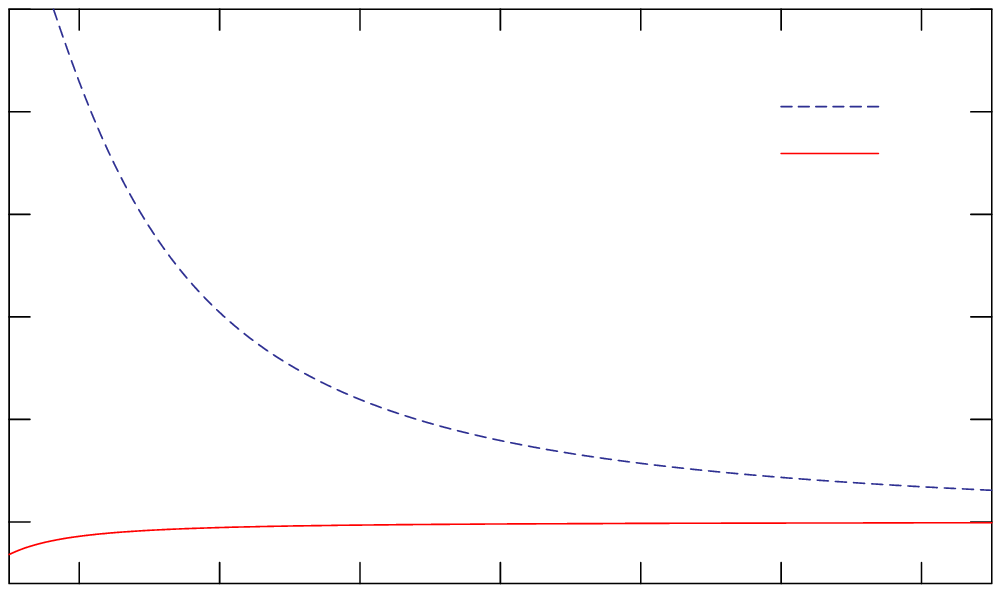}%
\end{picture}%
\setlength{\unitlength}{0.0200bp}%
\begin{picture}(18000,10800)(0,0)%
\put(2750,2866){\makebox(0,0)[r]{\strut{} 0}}%
\put(2750,4343){\makebox(0,0)[r]{\strut{} 1e-05}}%
\put(2750,5820){\makebox(0,0)[r]{\strut{} 2e-05}}%
\put(2750,7296){\makebox(0,0)[r]{\strut{} 3e-05}}%
\put(2750,8773){\makebox(0,0)[r]{\strut{} 4e-05}}%
\put(2750,10250){\makebox(0,0)[r]{\strut{} 5e-05}}%
\put(4036,1430){\makebox(0,0){\strut{} 0.02}}%
\put(6057,1430){\makebox(0,0){\strut{} 0.04}}%
\put(8079,1430){\makebox(0,0){\strut{} 0.06}}%
\put(10100,1430){\makebox(0,0){\strut{} 0.08}}%
\put(12121,1430){\makebox(0,0){\strut{} 0.1}}%
\put(14143,1430){\makebox(0,0){\strut{} 0.12}}%
\put(16164,1430){\makebox(0,0){\strut{} 0.14}}%
\put(550,5950){\rotatebox{90}{\makebox(0,0){\strut{}$D_2$}}}%
\put(10100,275){\makebox(0,0){\strut{}$E$ (GeV)}}%
\put(13868,8847){\makebox(0,0)[r]{\strut{}$5^{\circ}$}}%
\put(13868,8172){\makebox(0,0)[r]{\strut{}$90^{\circ}$}}%
\end{picture}
\caption{\it{$D_2$ for $\theta = 5^{\circ}$ (dashed line) and $90^{\circ}$
(continuous line).}}
\label{DIFF2l}
\ec
\end{figure}

In Figs.~\ref{DIFF1l} and~\ref{DIFF2l} we plotted (for a fixed value of the 
scattering angle) the quantities:
\bea
D_1 & = & \frac{\alpha}{\pi} \, \left( 
\left. \frac{d \sigma^{T}_1}{d \Omega} \right|_L - 
\frac{ d \sigma^{T}_1}{d \Omega} \right) \left(
\frac{d \sigma_0}{d \Omega} \right)^{-1} \, , \\
D_2 & = & \left( \frac{\alpha}{\pi} \right)^2 \,
\left( 
\left. \frac{d \sigma^{T}_2}{d \Omega} \right|_L -
\frac{ d \sigma^{T}_2}{d \Omega}\right) \left(
\frac{d \sigma_0}{d \Omega} + 
\frac{\alpha}{\pi} \frac{d \sigma_1^{T}}{d \Omega}  \right)^{-1} \, ,
\eea
respectively. A glance of the figures shows that the leading terms of the
cross-section approximate to a continuosly better degree the complete results
for increasing  beam energy. The leading terms of the cross-section fail to
reproduce the  complete result in the extremely forward and backward regions,
where $t$ or $u$  becomes smaller than $m^2$.

The codes for the numerical evaluation of the Bhabha scattering differential
cross-section in pure QED up to order $\alpha^4 (N_F =1)$ are available from 
the authors \cite{BHABHAPAGE}.

\section{Conclusions \label{Conc}}

In the present paper, we completed the evaluation of the Bhabha scattering 
cross-section at order $\alpha^4 (N_F =1)$ in pure QED. The calculation was
performed  without neglecting the electron mass $m$, and is valid for all
physical values of the independent Mandelstam invariants $s$ and $t$. The
master integrals necessary for the evaluation of the virtual corrections were
calculated in \cite{RoPieRem1} and \cite{us}, while the UV renormalized
unpolarized differential cross-section  was obtained in \cite{us2}.  The
calculation was completed by providing the real correction in the approximation
of a soft photon emission up to order $\alpha^4 (N_F =1)$, as well as by
explicitly showing that the IR poles present in such corrections cancel the
remaining IR poles of the virtual cross-section calculated in  \cite{us2}.
Finally, we developed computer codes for the numerical evaluation of the UV and
IR finite cross-section. We compared our findings with the results present in
the literature (where possible), obtaining complete agreement. We verified that
the effect of the terms proportional to positive powers of the electron mass
$m$ is negligible in the energy range of interest in present and future
colliders.

\section*{Acknowledgments}

We wish to thank S. Uccirati for very useful suggestions in the numerical
evaluation of our formulae and for allowing us to use his {\tt Fortran77} 
routines. 

We are grateful to J.~Vermaseren for his kind assistance in the use
of the algebra manipulating program {\tt FORM}~\cite{FORM}.

The work of R.~B. was supported by the European Union under
contract HPRN-CT-2000-00149 and by the 
Bundesministerium f\"ur Bildung und Forschung.
P.~M. has been supported by U.S. Department of Energy under grant
DE-FG03-91ER40662.

We would like to thank C.M.~Carloni~Calame for pointing out a sign mistake
in the one-loop vacuum polarization generating some sign mistakes in the 
formulas in appendices of the {\tt hep-ph/0405275v1} version of the present 
paper; these mistakes have been corrected in {\tt hep-ph/0405275v2}. Useful 
discussions with F.~Piccinini are also gratefully acknowledged \cite{CP}.

\appendix

\boldmath
\section{Integral $I_{i j}$ \label{app1}}
\unboldmath

As explained in Section~\ref{Real}, the integrals $I_{ij}$ are symmetric with
respect to their indices and satisfy the symmetry properties listed in
Eqs.~(\ref{Iijsym}). Therefore, in order to calculate the real corrections to
the Bhabha scattering cross-section up to the order that we consider in this
work, one needs to calculate the integrals $I_{1j}$ ($j =1,4$).  Following the
procedure outlined in Ref.~\cite{GP}, it is straightforward to verify that,
after the integration on the photon phase space has been carried out, the
integral becomes
\be
I_{1j} = \frac{\rho_{1j}}{2} \left[ \left(\frac{1}{D-4}  + \frac{1}{2} \ln\left(
\frac{\omega^2}{m^2}
\right) +
\ln{2} \right) I_{1j}^{(0)} - \Delta I_{1j}\right] + {\mathcal O} \left( D -4 \right)
\, . \label{Iijexp}
\ee
In the equation above, the quantity $\rho_{ij}$ has a particularly simple
expression in terms of the dimensionless quantities $x$, $y$, and $z$ introduced in
Eqs.~(\ref{eqx}-\ref{eqz}):
\be
\rho_{12} = \frac{1}{x} \, , \quad
\rho_{13} = \frac{1}{y} \, , \quad
\rho_{14} = \frac{1}{z} \, , \quad
\rho_{11} = 1\, . 
\ee

The quantity $I_{1j}^{(0)}$ in Eq.~(\ref{Iijexp}) can be expressed as
a simple integral as follows:
\be
I_{1j}^{(0)} = \frac{2}{ m^2} \int_0^1 dr \left[1 - 2 r \left(1 + \rho_{1j}
\frac{p_1 \cdot p_j }{m^2}
\right)  \right]^{-1} \,.
\ee
After observing that the scalar products $p_1 \cdot p_j$ also have
 a simple form
in terms of $x$, $y$, and $z$,
\be
p_1 \cdot p_2 = - \frac{m^2}{2}  \frac{1+x^2}{x} \, , \quad
p_1 \cdot p_3 = - \frac{m^2}{2}  \frac{1+y^2}{y} \, , \quad
p_1 \cdot p_4 = - \frac{m^2}{2}  \frac{1+z^2}{z} \, , \quad
p_1^2 = -m^2 \, ,
\ee
 one finds:
\bea
I_{11}^{(0)} &=& \frac{2}{m^2} \, , \\
I_{12}^{(0)} &=& -\frac{4}{m^2} \frac{x^2}{1-x^2} \ln{x} \, , \\
I_{13}^{(0)} &=& -\frac{4}{m^2} \frac{y^2}{1-y^2} \ln{y} \, , \\
I_{14}^{(0)} &=& -\frac{4}{m^2} \frac{z^2}{1-z^2} \ln{z}\, . 
\eea

In addition, the quantity $\Delta I_{1j}$ can also be expressed in integral form:
\be
\Delta I_{1j} = \int_0^1 dr \frac{1}{(P_{1j})^2} 
\frac{(P_{1j})_0}{|\vec{P_{1j}}|} 
\ln{\frac{(P_{1j})_0 - |\vec{P_{1j}}|}{(P_{1j})_0 + |\vec{P_{1j}}|}} \, , 
\label{Delta}
\ee
where the Lorentz vector $P_{1j}^\mu$ is defined by the relation
\be
P_{1j}^\mu = p_j^\mu + r \left( \rho_{1j} p_1^\mu - p_j^\mu \right) \, .
\ee
The integral in Eq.~(\ref{Delta}) can also be evaluated according to the
procedure outlined in Ref.~\cite{GP}; one finds:
\bea
\Delta I_{11} &=&  -\frac{1}{ m^2} \frac{1+x}{1-x} \ln{x} \, , \\
\Delta I_{1 l} &=& -\frac{2 }{ m^2 (\rho_{1 l}^2 -1 )} \Bigl[
\mbox{Li}_2 (a^{(1)}_l) + \mbox{Li}_2 (a^{(2)}_l) -
\mbox{Li}_2 (a^{(3)}_l) - \mbox{Li}_2 (a^{(4)}_l)
\Bigr] \, , \label{tt}
\eea
where $l = 2,3,4$ and the arguments $a^{(k)}_l$ ($k = 1,\ldots, 4$) have 
convenient expressions in terms of $x$ and $\rho_{1l}$:
\bea
a^{(1)}_l &=& \frac{1 - x \rho_{1l}}{1 + \rho_{1l}} \, , \quad\quad
a^{(2)}_l = \frac{x - \rho_{1l}}{x (1+ \rho_{1l})} \, , \\
a^{(3)}_l &=& \frac{\rho_{1l} - x}{1 + \rho_{1l}} \, , \quad\quad
a^{(4)}_l = -\frac{1 - x \rho_{1l}}{x (1+\rho_{1l}) } \, .
\eea
We observe that $a^{(1)}_2$ and $a^{(4)}_2$ are equal to zero; as a consequence,
the corresponding dilogarithms in Eq.~(\ref{tt}) also vanish.


\boldmath
\section{Leading Corrections at Order $\alpha^3$ and Order 
$\alpha^4(N_F=1)$ \label{expandedres}}
\unboldmath

In this Appendix, we provide the explicit expressions of the leading radiative
corrections defined in Eq.~(\ref{LL}).

%
\bea
\left. \frac{d \sigma_1^{T}}{d \Omega} \right|_L \! \!&=& \! \! \alpha^2
\Biggl\{
       - \ln\left(\frac{m^2}{s} \right)  \!\! \left(
           \frac{11}{3} \frac{t^2}{s^3}\! 
          + \! \frac{22}{3} \frac{t}{s^2} \! 
          + \! \frac{11}{s}\! 
          + \! \frac{11}{3} \frac{s}{t^2}\! 
          + \! \frac{22}{3 t}
          \right)
	- \ln^2\!\left( 
	  - \frac{t}{s}\right) \!\! \left(
	  \frac{t^2}{s^3}
          + \frac{13}{4} \frac{t}{s^2} \right. \nn \\
& &
       \left.
          + \frac{25}{4 s} 
          + 2 \frac{s}{t^2} 
          + \frac{19}{4 t}
          \right)  
	+   \ln\left( - \frac{t}{s}\right) \ln\left( 
	- \frac{u}{s}\right)   \left(
            2\frac{t^2}{s^3}   
          + 5\frac{t}{s^2}   
          +   \frac{19}{2 s}   
          +   4 \frac{s}{t^2}  
          +   \frac{8}{t}
          \right)  \nn \\
& &
       + \ln\left( \!- \frac{t}{s}\right)  \!\!  \left(\!
            \frac{11}{6} \frac{t}{s^2} \! 
          + \! \frac{11}{2 s}   
	  + \frac{11}{3}\frac{s}{t^2}
          + \!\frac{5}{t}\!
          \right)
      \!  - \ln^2\!\left(\!  - \frac{u}{s}\right) \! \! \left(
            \frac{t^2}{s^3}\! 
          + \!\frac{5}{2} \frac{t}{s^2}\! 
          + \! \frac{7}{2 s} \! 
          + \! \frac{5}{2 t} \!
          + \! \frac{s}{t^2} \!
          \right)\! \nn \\
& &
 	 + \! \ln\left( \!  - \frac{u}{s}\right)   \left(
           \frac{t}{2s^2} \! 
          + \! \frac{1}{2 t} 
          \right)
       + \! \zeta\left(2\right) \left( 2 \frac{t^2}{s^3} \! 
          + \! \frac{t}{s^2} \! 
          - \! \frac{9}{2s} \! 
          - 4 \frac{s}{t^2} 
          - \! \frac{8}{t}  \right)
	  - \frac{46}{9} \frac{t^2}{s^3}
          - \frac{92}{9} \frac{t}{s^2}\nn \\
& &
          - \frac{46}{3 s} 
          - \frac{46}{9} \frac{s}{t^2}
          - \frac{92}{9 t} 
- \left( \frac{t^2}{s^3}
          + 2 \frac{t}{s^2}
          + \frac{3}{s}
          + \frac{s}{t^2}
          + \frac{2}{t} \right) \Biggl[ 4\ln\left(2\right) \left(1 
	  + \ln\left( \frac{m^2}{s}\right)  \right. \nn \\
& &
        \left. 
	  +\ln\left( - \frac{u}{s} \right)
	  - \ln\left( - \frac{t}{s} \right) 
	  \right)
          + 4 \mbox{Li}_2\left(- \frac{u}{t}\right)   
          + 2 \ln\left(\frac{\omega^2}{s}\right)   \left( 1 
	     + \ln\left(\frac{m^2}{s}\right) 
	\right. \nn \\
& &
        \left.
	- \ln\left( - \frac{t}{s}\right) + \ln\left( 
	- \frac{u}{s}\right)\right)
	  \Biggr] \Biggr\} \, .
\eea

\bea
\left. \frac{d \sigma_2^{T}}{d \Omega} \right|_L \! \! &=& \! \!
-\alpha^2\Biggl\{
       \zeta\left(2\right)   \left(
           \frac{86}{9} \frac{t^2}{s^3}
          + \!\frac{263}{18} \frac{t}{s^2}
          + \!\frac{31}{3 s}
          + \!\frac{31}{9} \frac{s}{t^2}
          + \!\frac{61}{18t} 
          \right)
       - \zeta\left(3\right)   \left(
           2 \frac{t^2}{s^3}
          + 4 \frac{t}{s^2}
          + \!\frac{6}{s} \right. \nn  \\ 
& & \left.
          + 2 \frac{s}{t^2}
          + \frac{4}{t}
          \right)
	  + \mbox{Li}_2\left( - \frac{t}{s}\right) \ln\left( - \frac{t}{s}\right)   \left(
           \frac{2}{3} \frac{t^2}{s^3}
          + \frac{4}{3}\frac{t}{s^2}
          + \frac{2}{s}
          + \frac{2}{3} \frac{s}{t^2}
          + \frac{4}{3 t}
          \right)\nn  \\ 
& &	  
- \mbox{Li}_2\left( - \frac{u}{s}\right) \ln\left( - \frac{u}{s}\right)   \left(
           \frac{2}{3} \frac{t^2}{s^3}
          +\frac{t}{s^2}
          + \frac{1}{s}
          + \frac{1}{3 t}
          \right)
       - \mbox{Li}_2\left( -\frac{u}{t}\right) \ln\left(\frac{m^2}{s}\right)
       \times\nn  \\ 
& &	\times  
         \left(
           \frac{8}{3} \frac{t^2}{s^3}
          + \frac{16}{3}\frac{t}{s^2}
          + \frac{8}{s}
          + \frac{8}{3} \frac{s}{t^2}
          + \frac{16}{3 t}
          \right)
       + \mbox{Li}_2\left( -\frac{u}{t}\right) \ln\left( - \frac{t}{s}\right)   \left(
           \frac{5}{3}\frac{t}{s^2}
          + \frac{5}{s} \right. \nn  \\ 
& & \left.
          + \frac{10}{3} \frac{s}{t^2}
          + \frac{5}{t}
          \right)
       - \mbox{Li}_2\left( -\frac{u}{t}\right) \ln\left( - \frac{u}{s}\right)   \left(
           \frac{1}{3}\frac{t}{s^2}
          + \frac{1}{s}
          + \frac{2}{3} \frac{s}{t^2}
          + \frac{1}{t}
          \right)\nn  \\ 
& &
       - \mbox{Li}_2\left( -\frac{u}{t}\right)   \left(
           \frac{40}{9} \frac{t^2}{s^3}
          + \frac{80}{9}\frac{t}{s^2}
          + \frac{40}{3 s}
          + \frac{40}{9} \frac{s}{t^2}
          + \frac{80}{9 t}
          \right)
       - \mbox{Li}_3\left( - \frac{t}{s}\right)   \left(
           \frac{2}{3} \frac{t^2}{s^3}\right. \nn  \\ 
& & \left.
          + \frac{4}{3}\frac{t}{s^2}
          +\frac{2}{s}
          + \frac{2}{3} \frac{s}{t^2}
          + \frac{4}{3 t}
          \right)
       + \mbox{Li}_3\left( - \frac{u}{s}\right)   \left(
           \frac{2}{3} \frac{t^2}{s^3}
          +\frac{t}{s^2}
          + \frac{1}{s}
          + \frac{1}{3 t}
          \right)\nn  \\ 
& & 
       + \mbox{Li}_3\left( -\frac{t}{u}\right)   \left(
           \frac{1}{3}\frac{t}{s^2}
          + \frac{1}{s}
          + \frac{2}{3} \frac{s}{t^2}
          + \frac{1}{t}
          \right)
       + \!\ln\left(\frac{m^2}{s}\right) \zeta\left(2\right)   \left(
          \frac{8}{3}\frac{t^2}{s^3}
	  + \frac{7}{3}\frac{t}{s^2}
          - \frac{5}{2 s}\right. \nn  \\ 
& & \left.
          - \frac{10}{3} \frac{s}{t^2}
          - \frac{20}{3 t}
          \right)
       \!-\! \ln^3\!\left(\frac{m^2}{s}\right)  \! \!\left(
           \frac{1}{9} \frac{t^2}{s^3}
          + \frac{2}{9}\frac{t}{s^2}
          + \frac{1}{3 s}
          + \frac{1}{9} \frac{s}{t^2}
          + \frac{2}{9 t}
          \right)
       \!+\! \ln^2\!\left(\frac{m^2}{s}\right)\!\! \times  \nn  \\ 
& &\times
       \ln\left( - \frac{t}{s}\right)   \left(
          \frac{1}{3} \frac{t^2}{s^3}
          + \frac{2}{3}\frac{t}{s^2}
	  +\frac{1}{s}
          + \frac{1}{3} \frac{s}{t^2}
          + \frac{2}{3 t}
          \right)
       - \ln^2\left(\frac{m^2}{s}\right)
       \ln\left( - \frac{u}{s}\right)   \left(
           \frac{1}{3} \frac{t^2}{s^3} \right.\nn  \\ 
& & \left.
          + \!\frac{2}{3}\frac{t}{s^2}
          + \!\frac{1}{s}
          + \!\frac{s}{3 t^2}
          + \!\frac{2}{3 t}
          \right)
       - \ln^2 \left(\frac{m^2}{s}\right)\ln\left(\frac{\omega^2}{s}\right)   \left(
           \frac{4}{3} \frac{t^2}{s^3}
          + \!\frac{8}{3}\frac{t}{s^2}
          + \!\frac{4}{s}
          + \!\frac{4}{3} \frac{s}{t^2}\right.\nn  \\ 
& & \left.
          + \!\frac{8}{3 t}
          \right)
       - \ln^2\left(\frac{m^2}{s}\right)   \left(
           \frac{55}{18} \frac{t^2}{s^3}
          + \!\frac{55}{9}\frac{t}{s^2}
          + \!\frac{55}{6 s} 
          + \!\frac{55}{18} \frac{s}{t^2}
          + \!\frac{55}{9 t}
          \right)
       - \ln\left(\frac{m^2}{s}\right)\times\nn  \\ 
& & \times  \ln^2\left( - \frac{t}{s}\right)  \left(
           \frac{2}{3} \frac{t^2}{s^3}
          + \frac{31}{12}\frac{t}{s^2}
          + \frac{21}{4 s}
          + \frac{5}{3}\frac{s}{t^2}
          + \frac{49}{12 t}
          \right)
       + \ln\left(\frac{m^2}{s}\right) \ln\left( - \frac{t}{s}\right) \times\nn  \\ 
& & \times \ln\left( - \frac{u}{s}\right)   \left(
           \frac{4}{3} \frac{t^2}{s^3}
          + \frac{11}{3}\frac{t}{s^2}
          + \frac{15}{2 s}
          + \frac{10}{3} \frac{s}{t^2}
          + \frac{20}{3 t}
          \right)
       + \ln\left(\frac{m^2}{s}\right) \ln\left( - \frac{t}{s}\right) \times\nn  \\ 
& & \times \ln\left(\frac{\omega^2}{s}\right)   \left(
           \frac{4}{3} \frac{t^2}{s^3}
          + \!\frac{10}{3}\frac{t}{s^2}
          + \!\frac{6}{s}
          + \!\frac{8}{3} \frac{s}{t^2}
          + \!\frac{14}{3 t} 
          \right)
       + \!\ln\left(\frac{m^2}{s}\right) \ln\left( - \frac{t}{s}\right)  \times\nn  \\ 
& & \times  \left(
           \frac{10}{9} \frac{t^2}{s^3} \! 
          +\! \frac{85}{18}\frac{t}{s^2} \! 
          + \!\frac{65}{6 s} \! 
          + \!\frac{55}{9} \frac{s}{t^2} \! 
          + \!\frac{83}{9 t}
          \right)
      \! -\! \ln\left(\frac{m^2}{s}\right) \ln^2\left( - \frac{u}{s}\right)   \left(
           \frac{2}{3} \frac{t^2}{s^3}
         \! + \!\frac{11}{6}\frac{t}{s^2}\right.\nn  \\ 
& & \left. 
          + \!\frac{5}{2 s}\! 
          + \!\frac{2}{3} \frac{s}{t^2} \! 
          + \!\frac{11}{6 t}
          \right)
       - \ln\left(\frac{m^2}{s}\right) \ln\left( - \frac{u}{s}\right) \ln\left(\frac{\omega^2}{s}\right)   \left(
           \frac{4}{3} \frac{t^2}{s^3}\! 
          + \!\frac{8}{3}\frac{t}{s^2}\! 
          + \!\frac{4}{s}\! 
          + \!\frac{4}{3} \frac{s}{t^2}\right.\nn  \\ 
& & \left. 
          + \frac{8}{3 t}
          \right)
       - \ln\left(\frac{m^2}{s}\right) \ln\left( - \frac{u}{s}\right)   \left(
           \frac{10}{9} \frac{t^2}{s^3}
          + \frac{31}{18}\frac{t}{s^2}
          + \frac{10}{3 s}
          + \frac{10}{9} \frac{s}{t^2}
          + \frac{31}{18 t}
          \right)\nn  \\ 
& & 
       - \ln\left(\frac{m^2}{s}\right) \ln\left(\frac{\omega^2}{s}\right)   \left(
           \frac{32}{9} \frac{t^2}{s^3}
          + \frac{64}{9}\frac{t}{s^2}
          + \frac{32}{3 s}
          + \frac{32}{9} \frac{s}{t^2}
          + \frac{64}{9 t}
          \right)\nn  \\ 
& & 
       - \ln\left(\frac{m^2}{s}\right) \! \! \left(
           \frac{281}{27} \frac{t^2}{s^3} \! 
          + \! \frac{562}{27}\frac{t}{s^2}\! 
          + \! \frac{281}{9 s}\! 
          + \! \frac{281}{27} \frac{s}{t^2}
          + \! \frac{562}{27 t}
          \right)
       - \ln\left(\! - \frac{t}{s}\right) \!\zeta\left(2\right) \!  \left(
           2 \frac{t^2}{s^3}\right.\nn  \\ 
& & \left. 
          + \!\frac{5}{3}\frac{t}{s^2}
          - \!\frac{31}{6 s} 
          - \!4 \frac{s}{t^2}
          - \!\frac{9}{t}
          \right)
       - \ln^3\left(\! - \frac{t}{s}\right)   \left(
           \frac{1}{9} \frac{t^2}{s^3}
          - \!\frac{7}{9}\frac{t}{s^2}
          - \!\frac{28}{9 s}
          - \!\frac{16}{9} \frac{s}{t^2}
          - \!\frac{121}{36 t}
          \right)\nn  \\ 
& & 
       + \! \ln^2\left( - \frac{t}{s}\right) \ln\left( - \frac{u}{s}\right)  
          \left(
           \frac{1}{3} \frac{t^2}{s^3}
          - \frac{5}{6}\frac{t}{s^2}
          - \frac{25}{6 s}
          - \frac{10}{3} \frac{s}{t^2}
          - \frac{31}{6 t}
          \right)\! 
       - \! \ln^2 \left( - \frac{t}{s}\right)\times\nn  \\ 
& & \times \ln\left(\frac{\omega^2}{s}\right)   
       \left(
           \frac{2}{3}\frac{t}{s^2}\! 
          +\!\frac{2}{s}\! 
          +\! \frac{4}{3} \frac{s}{t^2}\! 
          +\! \frac{2}{t}
          \right)
       - \ln^2\! \left( - \frac{t}{s}\right) \! \! \left(
           \frac{10}{9} \frac{t^2}{s^3}\! 
          +\! \frac{44}{9}\frac{t}{s^2}\! 
          +\! \frac{34}{3 s}\! 
          +\! \frac{95}{18} \frac{s}{t^2}\right.\nn  \\ 
& & \left. 
          + \frac{86}{9 t}
          \right)
       - \ln\left( - \frac{t}{s}\right) \ln^2  \left( - \frac{u}{s}\right) \left(
           \frac{1}{3} \frac{t^2}{s^3}
          - \frac{13}{12 s}
          - \frac{s}{t^2}
          - \frac{3}{2 t}
          \right)
       + \ln\left( - \frac{t}{s}\right)\times\nn  \\ 
& & \times  \ln\left( - \frac{u}{s}\right) \ln\left(\frac{\omega^2}{s}\right)   \left(
           \frac{2}{3}\frac{t}{s^2} \! 
          +\! \frac{2}{s}\! 
          +\!  \frac{4}{3} \frac{s}{t^2}\! 
          + \! \frac{2}{t}
          \right)\! 
       + \! \ln\left( - \frac{t}{s}\right) \ln\left( - \frac{u}{s}\right)   \left(
           \frac{20}{9} \frac{t^2}{s^3}\right.\nn  \\ 
& & \left. 
          + \frac{55}{9}\frac{t}{s^2}\! 
          + \! \frac{77}{6 s}\! 
          + \! \frac{50}{9} \frac{s}{t^2}\! 
          + \! \frac{197}{18 t} 
          \right)
       + \ln\left( - \frac{t}{s}\right) \ln\left(\frac{\omega^2}{s}\right)   \left(
           \frac{20}{9} \frac{t^2}{s^3}
          + \frac{46}{9}\frac{t}{s^2}
          + \frac{26}{3 s}\right.\nn  \\ 
& & \left. 
          + \frac{32}{9} \frac{s}{t^2}
          + \frac{58}{9 t}
          \right)
       + \ln\left( - \frac{t}{s}\right)   \left(
           \frac{56}{27} \frac{t^2}{s^3}
          + \frac{449}{54}\frac{t}{s^2}
          + \frac{337}{18 s}
          + \frac{281}{27} \frac{s}{t^2}
          + \frac{418}{27 t}
          \right)\nn  \\ 
& &
       + \ln\left( - \frac{u}{s}\right) \zeta\left(2\right)   \left(
           2 \frac{t^2}{s^3}
          + \frac{19}{3}\frac{t}{s^2}
          + \frac{29}{3 s}
          + \frac{4}{3} \frac{s}{t^2}
          + \frac{17}{3 t}
          \right)
       - \ln^2 \left( - \frac{u}{s}\right)  \left(
           \frac{10}{9} \frac{t^2}{s^3}\right.\nn  \\ 
& & \left. 
          + \frac{29}{9}\frac{t}{s^2}
          + \frac{9}{2 s}
          + \frac{10}{9} \frac{s}{t^2}
          + \frac{29}{9 t}
          \right)
       + \ln^3\left( \! - \frac{u}{s}\right)   \left(
           \frac{1}{9} \frac{t^2}{s^3}
          + \frac{5}{18} \frac{t}{s^2}
          + \frac{5}{18s}
          + \frac{1}{6t}
          \right)
\nn  \\ & &
       - \ln\left(  \! - \frac{u}{s}\right) 
       \ln\left(\frac{\omega^2}{s}\right)   \left(
           \frac{20}{9} \frac{t^2}{s^3} \! 
          +  \! \frac{40}{9} \frac{t}{s^2} \! 
          +  \! \frac{20}{3s} \! 
          +  \! \frac{20}{9} \frac{s}{t^2} \! 
          +  \! \frac{40}{9t}
          \right)
\nn  \\ & &
       - \ln\left(  \! - \frac{u}{s}\right)   \left(
           \frac{56}{27} \frac{t^2}{s^3}
          + \frac{161}{54} \frac{t}{s^2}
          + \frac{56}{9s}
          + \frac{56}{27} \frac{s}{t^2}
          + \frac{161}{54t}
          \right)
       - \ln\left(\frac{\omega^2}{s}\right)   \left(
           \frac{20}{9} \frac{t^2}{s^3} \right. \nn  \\ 
& & \left.
          + \frac{40}{9} \frac{t}{s^2}
          + \frac{20}{3s}
          + \frac{20}{9} \frac{s}{t^2}
          + \frac{40}{9t}
          \right)
       - \ln\left(2\right) \ln^2 \left(\frac{m^2}{s}\right)  \left(
           \frac{8}{3} \frac{t^2}{s^3}
          + \frac{16}{3} \frac{t}{s^2}
          +  \frac{8}{s} \right. \nn  \\ 
& & \left.
          + \frac{8}{3} \frac{s}{t^2}
          + \frac{16}{3t}
          \right)
       + \ln\left(2\right) \ln\left(\frac{m^2}{s}\right) \ln\left( - \frac{t}{s}\right)   \left(
           \frac{8}{3} \frac{t^2}{s^3}
          + \frac{20}{3} \frac{t}{s^2}
          + \frac{12}{s}
          + \frac{16}{3} \frac{s}{t^2} \right. \nn  \\ 
& & \left.
          + \frac{28}{3t}
          \right)
       - \ln\left(2\right) \ln\left(\frac{m^2}{s}\right) \ln\left( - \frac{u}{s}\right)   \left(
           \frac{8}{3} \frac{t^2}{s^3}
          + \frac{16}{3} \frac{t}{s^2}
          + \frac{8}{s}
          + \frac{8}{3} \frac{s}{t^2}
          + \frac{16}{3t} 
          \right)
\nn  \\ & &
       - \ln\left(2\right) \ln\left(\frac{m^2}{s}\right)   \left(
           \frac{64}{9} \frac{t^2}{s^3}
          + \frac{128}{9} \frac{t}{s^2}
          + \frac{64}{3s} 
          + \frac{64}{9} \frac{s}{t^2}
          + \frac{128}{9t}
          \right)
\nn  \\ & &
       - \ln\left(2\right) 
         \ln^2\left( \! - \frac{t}{s}\right)   \left(
           \frac{4}{3} \frac{t}{s^2} \! 
          + \!  \frac{4}{s} \! 
          +  \! \frac{8}{3} \frac{s}{t^2} \! 
          +  \! \frac{4}{t}
          \right) \! 
       +  \! \ln\left(2\right) \ln\left( \!  - \frac{t}{s}\right) 
       \ln\left(  \! - \frac{u}{s}\right)   \left(
           \frac{4}{3} \frac{t}{s^2} \right. \nn  \\ 
& & \left.
          + \frac{4}{s}
          + \frac{8}{3} \frac{s}{t^2}
          + \frac{4}{t}
          \right)
       + \ln\left(2\right) \ln\left( - \frac{t}{s}\right)   \left(
           \frac{40}{9} \frac{t^2}{s^3}
          + \frac{92}{9} \frac{t}{s^2}
          + \frac{52}{3s}
          + \frac{64}{9} \frac{s}{t^2} \right. \nn  \\ 
& & \left.
          + \frac{116}{9t} 
          \right)
       - \ln\left(2\right) \ln\left( - \frac{u}{s}\right)   \left(
           \frac{40}{9} \frac{t^2}{s^3}
          + \frac{80}{9} \frac{t}{s^2}
          + \frac{40}{3s} 
          + \frac{40}{9} \frac{s}{t^2}
          + \frac{80}{9t}
          \right)
\nn  \\ & &
       - \ln\left(2\right)   \left(
           \frac{40}{9} \frac{t^2}{s^3}
          + \frac{80}{9} \frac{t}{s^2}
          + \frac{40}{3s} 
          + \frac{40}{9} \frac{s}{t^2}
          + \frac{80}{9t} 
          \right)
       - \frac{1967}{108} \frac{t^2}{s^3}
          - \frac{1967}{54} \frac{t}{s^2} \nn  \\ 
& & 
          - \frac{1967}{36s}
          - \frac{1967}{108} \frac{s}{t^2}
          - \frac{1967}{54t} \Biggr\}\, .
\eea


\end{document}